\documentclass[twocolumn,prl,10pt,aps,superscriptaddress,longbibliography]{revtex4-1}

\usepackage[english]{babel}
\PassOptionsToPackage{English}{babel}
\usepackage{amsmath,revsymb,graphicx,amssymb,array,bm,adjustbox,multirow,soul,changes,changebar}

\newcommand{\bbeta}{\bm{\beta}}
\newcommand{\bgamma}{\bm{\gamma}}
\newcommand{\COM}{\mathrm{COM}}
\newcommand{\rel}{\mathrm{rel}}

\begin{document}

\title{Integrable families of hard-core particles with unequal masses in a one-dimensional harmonic trap}

\author{N.L. Harshman}\email{Electronic address: harshman@american.edu}
\affiliation{Department of Physics, American University, 4400 Massachusetts Ave.\ NW, Washington, DC 20016, USA}
\affiliation{Department of Physics and Astronomy, Aarhus University, DK-8000 Aarhus C, Denmark}
\author{Maxim Olshanii}
\affiliation{Department of Physics, University of Massachusetts Boston, Boston, MA 02125, USA}
\author{A.S. Dehkharghani}
\affiliation{Department of Physics and Astronomy, Aarhus University, DK-8000 Aarhus C, Denmark}
\author{A.G. Volosniev}
\affiliation{Institut f\"ur Kernphysik, Technische Universit\"at Darmstadt, 64289 Darmstadt, Germany}
\author{Steven Glenn Jackson}
\affiliation{Department of Mathematics, University of Massachusetts Boston, Boston, MA 02125, USA}
\author{N.T.~Zinner}
\affiliation{Department of Physics and Astronomy, Aarhus University, DK-8000 Aarhus C, Denmark}

\begin{abstract}
We show that the dynamics of particles in a one-dimensional harmonic trap with hard-core interactions can be solvable for certain arrangements of unequal masses. For any number of particles, there exist two families of unequal mass particles that have integrable dynamics, and there are additional exceptional cases for three, four and five particles. The integrable mass families are classified by Coxeter reflection groups and the corresponding solutions are Bethe ansatz-like superpositions of hyperspherical harmonics in the relative hyperangular coordinates that are then restricted to sectors of fixed particle order. We also provide evidence for superintegrability of these Coxeter mass families and conjecture maximal superintegrability.
\end{abstract}

\maketitle

The complexity of interacting quantum systems can be partially tamed by extrapolating from solvable models, especially in one dimension~\cite{albeverio_solvable_2012, sutherland_beautiful_2004, gaudin_bethe_2014}. Prominent examples include the Lieb-Liniger model with zero-range contact interactions in free space~\cite{lieb_exact_1963}, the Tonks-Girardeau gas with hard-core contact interactions~\cite{girardeau_relationship_1960}, the Calogero-Moser (CM) model with inverse square interactions either free or in a harmonic trap~\cite{calogero_solution_1971, calogero_calogero-moser_2008}, and the extended family of Calogero-Sutherland-Moser (CSM) models~\cite{olshanetsky_quantum_1983}. Such models provide insights about the dynamics and thermodynamics of few-body and many-body physics, and they are proving grounds for inquiry into the nature of integrability, solvability, and chaos.

Interest in one-dimensional models has surged because of experiments with ultracold atoms trapped in tight wave guides with interactions controlled by Feshbach and confinement-induced resonances~\cite{cazalilla_one_2011, guan_fermi_2013}. These systems are well-described by a one-dimensional model with contact interactions~\cite{olshanii_atomic_1998}. Dynamical effects predicted by this model like delayed thermalization due to integrability at the hard-core limit have been observed~\cite{kinoshita_quantum_2006}. Controllable dynamics and extended coherence times, possibly combined with internal degrees of freedom like spin or hyperfine structure, make such atomic systems suitable for exploring fundamental few-body and many-body quantum physics~\cite{serwane_deterministic_2011, wenz_few_2013} as well as for applications in quantum technologies~\cite{divincenzo_universal_2000, PhysRevLett.91.207901, PhysRevA.91.023620}. 

However, the famous models mentioned above primarily consider equal-mass particles~\footnote{An exception is the free-space CSM model with mass-scaled interaction strengths in Ref.~\cite{sen_multispecies_1996}.}. This article analyzes one-dimensional particles with different masses (but the same frequencies) in a harmonic trap with hard-core contact interactions. Our analysis shows that for particles with certain masses in a certain order, the mass-imbalanced hard-core system is integrable. Conversely, we provide numerical evidence that for other masses, or even the same masses but in a different order, the dynamics are quantum ergodic. Both of these limits possess potentially observable signatures in the energy level statistics~\cite{bohigas1991, whan_hierarchical_1997}, particle correlations~\cite{PhysRevA.89.053623}, and thermalization dynamics~\cite{rigol_thermalization_2008, PhysRevA.89.033601}. Measurements probing the relationship between ergodicity and entanglement in a closed quantum system have recently been performed in superconducting qubits~\cite{neill_ergodic_2016}, showing the power for controllable quantum systems to test non-equilibrium thermodynamics.

The possibility for experimental implementation of mass-imbalanced atomic systems has driven multiple recent analyses. For general masses with contact interactions in an equal-frequency harmonic trap there are no exact solutions, so most previous approaches have relied on approximation schemes and numerical methods~\cite{mehta_born-oppenheimer_2014, dehkharghani_quantum_2015, pecak_two-flavour_2016, mehta_few-boson_2015, PhysRevA.94.042118, dehkharghani_impenetrable_2016, pecak_experimentally_2017, PhysRevA.95.053632}. In this article, we show that for hard-core interactions there exist families of unequal masses for which there are exact solutions for the ground state and all excited states. This extends results first derived for hard-core interactions in free space~\cite{olshanii_exactly_2015, olshanii_exactly_2016, olshanii_creating_2016}. Exact solutions for hard-core interactions form the basis for approximation schemes for strongly-interacting systems, providing a valuable benchmark for testing numerical methods~\cite{volosniev_strongly_2014, PhysRevA.90.013611, Harshman2016}.  This possibility is of special importance for mass imbalanced systems where \emph{ab initio} calculations for strong interaction are especially challenging~\cite{0954-3899-41-5-055110}.

We derive the criteria for which sets of imbalanced masses are solvable and integrable using a geometrical approach. In the hard-core limit, configuration space is sectioned into $N!$ disconnected sectors, one for each order of particles. After separating out the center-of-mass and relative hyperradial degrees of freedom, the dynamics in each ordering sector of relative angular configuration space can be mapped onto hard-wall quantum billiards on a $(N\!-\!2)$-dimensional sphere. The domain is a simplex whose shape depends on the mass ratios. For three particles, the six ordering sectors are just arcs of a circles and every set of masses is therefore integrable by separation of variables. For four particles, the 24 ordering sectors are spherical triangles. Quantum billiards in a general spherical triangle cannot be solved by separability, nor can higher dimensional generalizations to $(N\!-\!2)$-simplexes on $(N\!-\!2)$-spheres. However, when a particular ordering sector tiles the sphere under reflections, then the problem is exactly solvable using something like the method of images. The possible spherical tilings are classified using Coxeter groups. Described in more detail below, Coxeter groups are point symmetry groups generated by reflections. They were originally developed for the purposes of analyzing symmetric polytopes~\cite{coxeter_regular_1973}. In hard-core contact interaction models, the $(N\!-\!1)$-dimensional hyperplanes where two particles coincide define planes of reflection. If the particle masses are correct, then these coincidence reflection planes generate a Coxeter group.

This logic can be reversed: we show that for every finite, connected, non-branching Coxeter group with rank $r$, there is one-parameter family of masses for which the dynamics of $r+1$ ordered particles is integrable. These models are non-trivial when $N > 3$; we focus on the case $N=4$ where there are three families of solvable masses and the symmetries in relative configuration space are the same as the Platonic solids. We give special attention to the exceptional Coxeter group $H_3$ of icosahedral symmetries, and therefore this work is closely related to CSM models based on exceptional reflection groups~\cite{boreskov_solvability_2005, garcia_quantum_2010, garcia_quantum_2011}.  The role of Coxeter groups in providing integrability criteria for this model is perhaps not surprising because they have previously played an important role in the theory of classical and quantum dynamical systems. For example, extensions of the CSM have a closely-related classification scheme~\cite{olshanetsky_quantum_1983}, and so do Gaudin models~\cite{gaudin_bethe_2014}.

Besides applications to mass-imbalanced ultracold atomic gases, a motivating interest in this model is that it sits at the intersection of related notions of integrability and solvability. For sectors with Coxeter tiling symmetry, the energies can be calculated algebraically and all excited states can be expressed as orthogonal polynomials times the ground state; this property is called \emph{exact solvability}~\cite{tempesta_exact_2001, post_families_2012}. These solutions are constructed by Bethe-ansatz like superpositions, not of plane waves, but of spherical (or hyperspherical) harmonics. The conditions on the masses can be seen as the requirement that the scattering is non-diffractive, i.e.\ integrable in the Bethe-ansatz sense~\cite{mcguire_study_1964, sutherland_nondiffractive_1980, sutherland_beautiful_2004, lamacraft_diffractive_2013}. On the other hand, we provide evidence that these models are also integrable in the classical, Liouvillian sense. Classically, Liouvillian integrability means there are $N$ functionally-independent invariant operators in involution that act continuously on $2N$ dimensional phase space. The quantum version of these operators commute with each other and therefore all states can be characterized by the spectrum of this set of operators (for a discussion of ambiguities defining integrability in quantum systems, see~\cite{caux_remarks_2011}). As an example, we construct these operators for the four-particle case of $H_3$. Further, we conjecture that these solvable models are superintegrable, meaning they have more integrals of motion than degrees of freedom, and even  \emph{maximally superintegrable} with $2N-1$ integrals of motion (for a more complete discussion, see~\cite{evans_superintegrability_1990}). Superintegrable systems can have rich mathematical structure, like multi-separability and exact solvability. Further, perturbations from superintegrable models sometimes remain integrable. For example, the unitary (hard-core) limit of equal-mass particles in a harmonic trap with contact interactions is a maximally superintegrable system isomorphic to one limit of the CM model~\cite{wojciechowski_superintegrability_1983, calogero_calogero-moser_2008, hakobyan_superintegrability_2014}. In the near-unitary limit, defined as a first order perturbation from the hard-core limit, maximal superintegrability is broken. However, the system still retains enough symmetry in the near-unitary limit that it can be mapped onto a spin chain model~\cite{volosniev_strongly_2014, PhysRevA.90.013611}. At the other extreme from maximal superintegrability, the Coxeter group criteria could be used to identify mass-imbalanced systems where quantum ergodic dynamics should be expected. In the experimental outlook, we discuss the connections to quantum billiards and the consequences of integrability for thermalization in ultracold atomic gases.

\section*{Model and Symmetry}

We consider the $N$-particle Hamiltonian with contact interactions. All particles are harmonically trapped with the same frequency. Written in terms of the particle coordinates ${\bf x} = (x_1, x_2, \ldots, x_N)$, the Hamiltonian has the form
\begin{equation}\label{eq:hamx}
H = \sum_{i=1}^N {\left(\! \frac{-\hbar^2}{2 m_i} \frac{\partial^2}{\partial x_i^2} +\frac{1}{2}m_i \omega^2 x_i^2 \! \right)} +   g \sum_{i<j} \delta( x_i - x_j).
\end{equation}
In the limit $g \to \infty$ of hard-core contact interactions, the order of the particles is a dynamical invariant~\cite{Harshman2016}. Configuration space is divided into $N!$ ordering sectors by $N(N-1)/2$ impenetrable coincidence hyperplanes, one for each pair of particles. The order $x_{p_1}\! \leq\! x_{p_2}\! \leq \!\cdots\! \leq\! x_{p_N}$ is labeled by a permutation $p=\{p_1, p_2, \ldots, p_N\}$, or more briefly $p=p_1p_2\cdots p_N$.   Denote by $X_{ij}$ the coincidence hyperplane defined by $x_i-x_j=0$.  

\begin{figure}
\adjustbox{trim={.29\width} {.28\height} {0.29\width} {.25\height},clip}{\includegraphics[width= 2\columnwidth]{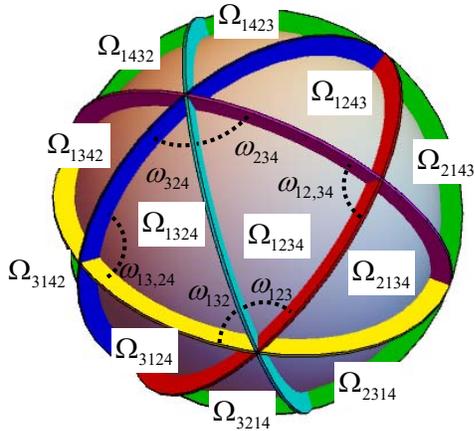}}
\caption{This figure depicts the relative configuration space for four particles in the simplest case when all masses are the same, an example of the one-parameter mass family of the Coxeter group $A_3$. The gray sphere represents an equipotential of the harmonic trap in the mass-normalized coordinates $(z_1,z_2,z_3)$. The six colored disks that intersect the plane represent the coincidence planes $Z_{12}$ (red), $Z_{13}$ (yellow), $Z_{14}$ (green), $Z_{23}$ (cyan), $Z_{24}$ (blue), $Z_{34}$ (magenta). Twelve sectors are visible and are labeled by $\Omega_p$, where $p$ is the order of the four particles. For two sectors $\Omega_{1234}$ and $\Omega_{1324}$, the three angles are also labeled.}\label{fig:AAA}
\end{figure}

In Appendix A, we show that solving for the eigenstates of (\ref{eq:hamx}) in a particular ordering sector $p$ is equivalent to solving for the motion of a free quantum particle confined to an $(N\!-\!2)$-sphere and trapped inside an angular sector $\Omega_p$ bounded by $(N\!-\!1)$ hard walls. We establish this equivalence by making a mass-dependent transformation $T$ of the position coordinates ${\bf z} = T {\bf x}$, and then separating out the scaled center-of-mass $z_N$ and the scaled relative hyperradius $\rho$
\begin{equation}
z_N = \sqrt{\frac{\omega }{\hbar M}} \sum_{i=1}^N m_i x_i,\ \mbox{and}\ 
\rho^2 = \sum_{j=1}^{N-1} z_j^2,
\end{equation}
where $M$ is the total mass. The remaining relative coordinates are the $(N\!-\!2)$ hyperangles $\{\phi,\theta_1,\ldots,\theta_{N\!-\!3}\}$ that cover the sphere $\mathcal{S}^{N\!-\!2}$. The transformation to these coordinates gives the harmonic potential a spherically symmetric form, but the coincidence hyperplanes, now transformed to $Z_{ij} = T (X_{ij})$, break that symmetry. For later convenience, denote by $\hat{\bgamma}_{ij}$ the unit normal vector to the $Z_{ij}$ coincidence hyperplane.

The specific angular ordering sector $\Omega_p$ is bounded by the intersection of the sphere with the $(N\!-\!1)$ coincidence hyperplanes $Z_{p_1 p_{2}}$, $Z_{p_2 p_3}$, \ldots, $Z_{p_{N\!-\!1} p_N}$.Each sector $\Omega_p$ has $(N\!-\!2)$ angles $\omega_{ijk}$ of the form
\begin{equation}\label{eq:angle}
\omega_{ijk} = \arctan\left(\sqrt{\frac{m_j (m_i + m_j + m_k)}{m_i m_k}}\right).
\end{equation}
corresponding to the intersections of coincidence planes $Z_{ij}$ and $Z_{jk}$ that share a particle, and $(N\!-\!3)(N\!-\!2)/2$ angles of $\omega_{ij,kl}=\pi/2$ for the intersections of coincidence planes $Z_{ij}$ and $Z_{kl}$ that do not share a particle. For four particles each ordering sector $\Omega_p=\Omega_{ijkl}$ is a spherical triangle bounded by three great circles; see Fig.~\ref{fig:AAA}.

Solving the (hyper)spherical Helmholtz equation on an angular sector $\Omega_p$ with Dirichlet boundary conditions is an example of quantum billiards. The problem of quantum and classical billiards in planar triangles is well-studied~\cite{casati_computer_1976, richens_pseudointegrable_1981, gutkin_billiards_1996, glashow_three_1997, artuso_numerical_1997, casati_mixing_1999, mcguire_extending_2001, araujo_lima_ergodicity_2013, wang_nonergodicity_2014}, and the integrability and solvability of the dynamics depends critically on the domain shape of the billiards. For example, the only three triangular billiards in a plane that have classically-integrable dynamics are the three triangles with distinguishable sides that tile the plane under reflections, without gaps or overlaps (see footnote 3 of \cite{olshanii_exactly_2015}). This serves as our guide for the following result for spherical quantum billiards. \emph{The dynamics in an angular sector $\Omega_p$ is integrable and exactly solvable when:}
\begin{itemize}
\item The sector $\Omega_p$ tiles the $(N\!-\!2)$-sphere under reflections across its boundaries. The tiling covers the sphere with no gaps or overlaps and distinguishable sides. In other words, the $(N\!-\!2)(N\!-\!1)/2$ angles of a sector $\omega_{ijk}$ and $\omega_{ij,kl}$ define a spherical kaleidoscope.
\item The $(N\!-\!1)$ reflections across the bounding hyperplanes $Z_{p_1p_2}$, $Z_{p_2p_3}$, \ldots, $Z_{p_{N-1}p_N}$ generate a finite Coxeter reflection group. The $(N\!-\!1)$ reflection normals $\hat{\bgamma}_{p_1p_2}$, $\hat{\bgamma}_{p_2p_3}$, \ldots, $\hat{\bgamma}_{p_{N-1}p_N}$ are the simple roots of the Coxeter group. 
\end{itemize}

All finite reflection groups (in all dimensions) were classified by Coxeter~\cite{coxeter_regular_1973, humphreys_reflection_1992}. Abstractly, a Coxeter group of rank $m$ is a finite group generated by $m$ reflections, where a reflection is a group element that squares to the identity. Every point symmetry group in $m$ dimensions is either a Coxeter group or a subgroup of a Coxeter group of rank $m$. For example, the three-dimensional point groups familiar from chemical and solid state physics are all subgroups of the the Coxeter groups $A_3$ (tetrahedral symmetry), $C_3$ (cubic symmetry), and $H_3$ (icosahedral symmetry, or they are subgroups of products of lower rank Coxeter groups.

The structure of the reflection group can be encoded by the Coxeter diagram, which can be branching or non-branching and connected or not connected. There is a family of $N$ masses that determine a `good' sector for every \emph{non-branching} and \emph{connected} Coxeter reflection group with rank $N\!-\!1$. These groups are listed in Table \ref{tab:coxeter}.  Only the non-branching Coxeter groups are relevant because in one-dimension each pair can have at most two adjacent pairs. Geometrically, this enforces that the coincidence planes of non-adjacent pairs like $Z_{ij}$ and $Z_{kl}$ are orthogonal $\omega_{ij,kl} = \pi/2$. We focus our attention on the connected Coxeter groups because they are relevant when all masses are finite. Disconnected graphs realize limiting cases of extreme mass imbalances. For example, for four particles, if the first or fourth particle is much more massive, i.e.\ like the `Born-Oppenheimer' case of \cite{mehta_born-oppenheimer_2014}, then the reducible Coxeter groups like $I_2(q) \times {A}_1$ could be employed.

\begin{table}
\caption{This table provides information about the connected, non-branching, finite Coxeter reflection groups. For $N$ particles, each rank $m=N-1$ Coxeter group defines a one-parameter family of masses for which the system is exactly solvable. For each group $G_m$, the following data are provided~\cite{goodman_alice_2004}: the Coxeter bracket $[q_1,\ldots,q_m]$ from which one determines the angles of the integrable sector; the number of reflections $\lambda_0$ in the group which determines the relative angular momentum of the ground state solution; and the order $G$ of the group which gives the number of integrable sectors required to tile the sphere. Note that there are two series of groups $A_m$ and $C_m$ that provide integrable mass families for any number of particles.}
\centering
\label{tab:coxeter}
\begin{tabular}{c|c|c|c|c}
\hline
$N$ & $A$-series & $C$-series & $H$-type & others \\
\hline
\multirow{4}{*}{$3$}& $A_2 \equiv I_2(3) $ & $C_2 \equiv I_2(4) $ & $H_2 \equiv I_2(5) $ & $I_2(q) $\\
&$[3]$ & $[4]$ & $[5]$ & $[q]$ \\
&$\lambda_0=3$ & 	$\lambda_0=4$ &  $\lambda_0=5$ & $\lambda_0=q$ \\
&$G=6$ & 	$G=8$ &  $G=10$ & $G=2q$ \\
\hline
\multirow{4}{*}{$4$}& $A_3 $ & $C_3  $ & $H_3  $ &  \\
&$[3,3]$ & $[4,3]$ & $[5,3]$ & \\
& $\lambda_0=6$ & 	$\lambda_0=9$ & $\lambda_0=15$ &  \\
& $G=24$ & 	$G=48$ & $G=120$ &  \\
\hline
\multirow{4}{*}{$5$}& $A_4 $ & $C_4  $ & $H_4  $ & $F_4$ \\
&$[3,3,3]$ & $[4,3,3]$ & $[5,3,3]$ & $[3,4,3]$\\
& $\lambda_0=10$ & 	$\lambda_0=16$ & $\lambda_0=60$ & $\lambda_0=24$ \\
& $G=120$ & 	$G=384$ & $G=14400$ & $G=1152$ \\
\hline
\multirow{4}{*}{$ \geq\! 6$}& $A_{N\!-\!1} $ & $C_{N\!-\!1}  $ &  &  \\
&$[3^{N-1}]$ & $[4,3^{N-2}]$ &  & \\
& $\lambda_0=\frac{N (N\!-\!1)}{2}$ & 	$\lambda_0=(N\!-\! 1)^2$ &  &  \\
& $G=(N\!-\!1)!$ & 	$G=2^N N!$ &  &  \\
\hline

\end{tabular}

\end{table}

\begin{figure}
\adjustbox{trim={.30\width} {.2\height} {0.30\width} {.16\height},clip}{\includegraphics[width= 2\columnwidth]{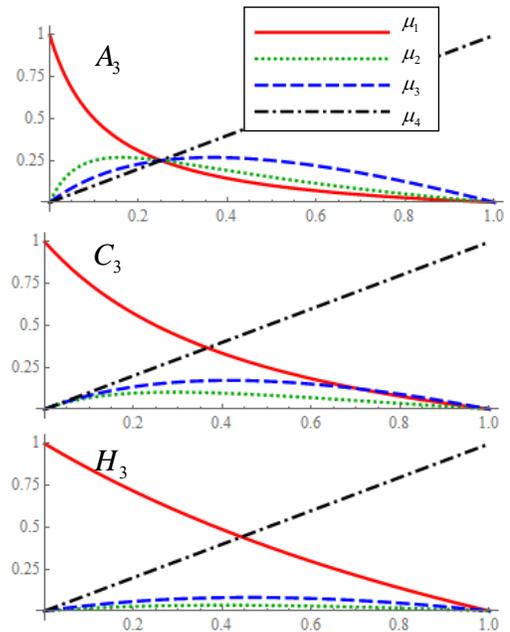}}
\caption{This figure depicts the integrable mass families for four particles with the Coxeter symmetries $A_3$, $C_3$, and $H_3$. The mass fractions $\mu_i=m_i/M$  are plotted versus $\mu_4$ (legend is in top graph). The case where all masses are the same $\mu_i = 1/4$ is in mass family $A_3$. Mass family $C_3$ includes two cases where two finite masses are the same, and $H_3$ includes one.}\label{fig:CCC}
\end{figure}

For each group in Table \ref{tab:coxeter}, there exists a one-parameter family of mass sequences for which there are integrable sectors. The bracket notation for the Coxeter group $[q_1,q_2,\ldots,q_{N-2}]$ determines the sector angle by eq.\ \ref{eq:angle}, where $\omega_{i (i+1) (i+2)} = \pi/q_i$. In Figure \ref{fig:CCC}, we show the integrable mass spectra for the three four-particle families $A_3$, $C_3$, and $H_3$. This can be reversed: given a set of $N$ masses in a particular order, one could check how close the sector angles derived from the masses come to the angles $\pi/q_1, \ldots, \pi/q_{N-2}$ that define a rank $N-1$ Coxeter group.

\begin{figure}
\adjustbox{trim={.33\width} {.33\height} {0.24\width} {.31\height},clip}{\includegraphics[width= 2\columnwidth]{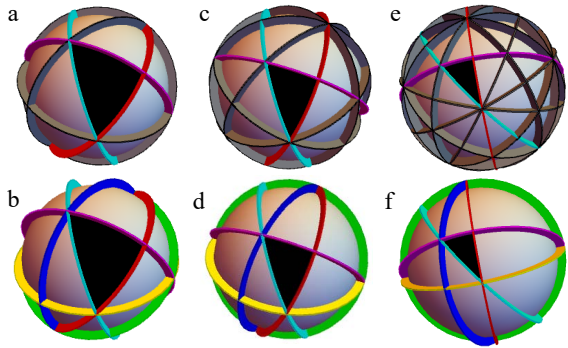}}
\caption{The top row of figures (a, c, e) depict the arrangement of the $\lambda_0$ reflection planes (gray and colored disks) and the tiling of sphere into $G$ spherical triangle sectors for $A_3$, $C_3$ and $H_3$, respectively. The bottom row of figures (b, d, f) show the coincidence planes (colored disks) for specific, non-symmetric choices of masses within the mass families for $A_3$, $C_3$ and $H_3$, respectively. The disk colors are the same as in Fig.~1. The black spherical triangle tile the top figure and is similar to the integrable sector $\Omega_{1234}$ in the bottom figure. This sector is bounded by the planes $Z_{12}$ (red), $Z_{23}$ (cyan), $Z_{34}$ (magenta), which are the generating planes for the Coxeter symmetry.}\label{fig:DDD}
\end{figure}

For $N$ particles that define a `good' sector, one that tiles the $(N\!-\!2)$-sphere, the generators of the Coxeter group can be chosen as the $m=N-1$ reflections $R_{ij}$ across the boundary hyperplanes $Z_{ij}$ of the sector $p={12\ldots N}$. 
For $N=4$, the Coxeter groups are rank $m=3$ and are generated by $R_{12}$, $R_{23}$, and $R_{34}$. All three generators square to the identity and the relations 
\begin{equation}
(R_{12}R_{34})^2 = (R_{34}R_{23})^3 =  (R_{23}R_{12})^q =1,
\end{equation}
hold for $q=3$, $q=4$, and $q=5$ for $A_3$, $C_3$ and $H_3$, respectively. Generally, within each Coxeter group $G_m$, there is a conjugacy class $K \subset G_m$ of all reflections $R\in G_m$. We denote the number of reflection planes (and the order of $K$) by ${\lambda_0}$ and denote the normals to these planes by $\hat{\bgamma}(R)$, but remember that only $N\!-\!1$ of these planes and normals are ``real'', i.e.\ they correspond to actual coincidence planes and normals. See Figure \ref{fig:DDD}.

Note that for the Coxeter groups in the $A$-series, $C$-series and the exceptional group $F_4$, the sector angles are all $\pi/2$, $\pi/3$ or $\pi/4$. Inspecting eq.~\ref{eq:angle}, we see that for these cases the masses are all rational fractions of each other. For example, for the group $C_3$ there are an countably-infinite number of rational mass sequences that give integrable sectors. The four with the lowest rational denominators are given by $(3m,m,2m,6m)$, $(10m,2m,3m,5m)$, $(12m,3m,5m,10m)$, and $(56m,7m,9m,12m)$. As discussed below, this allows the possibility of building integrable systems out of clusters of particles with the same mass.

\section*{Exact solvability and Bethe-ansatz integrability}

For each Coxeter group, there is therefore a one-parameter family of masses such that the complete spectrum of energy eigenstates can be exactly solved in the ordering sector $\Omega_{1\cdots N}$ and its inverted sector $\Omega_{N\cdots 1}$. The ground state in each of these `Coxeter sectors' is non-degenerate and its hyperangular wave function can be expressed as
\begin{equation}\label{eq:Ups0}
\Upsilon_{\lambda_0}(\hat{\bf z}) = 
N_{\lambda_0} \prod_{R \in K}{(\hat{\bgamma}(R) \cdot \hat{\bf z}) },
\end{equation}
where $\hat{\bf z} = (z_1,\ldots, z_{N-1})/\rho$ is a unit vector expressed in hyperspherical coordinates and $N_{\lambda_0}$ is a normalizing factor. For all $R\in K$, the function (\ref{eq:Ups0}) is reflection antisymmetric $\Upsilon_{\lambda_0}(R \hat{\bf z})= -\Upsilon_{\lambda_0}(\hat{\bf z})$ and therefore vanishes on all reflections planes, including the coincidence planes. Note that $\rho^{\lambda_0}\Upsilon_{\lambda_0}(\hat{\bf z})$ is the lowest degree anti-invariant polynomial of the corresponding group \cite{humphreys_reflection_1992}. 
The function (\ref{eq:Ups0}) is defined on the entire sphere, but its restriction to the ordering sectors $\Omega_{1\cdots N}$ or $\Omega_{N\cdots 1}$ provides that sector's ground state with energy $\hbar\omega({\lambda_0} + N/2)$. Exploiting separability, a tower of states are laddered from the ground state manifold with energies $\hbar\omega(n + 2\nu + {\lambda_0} + N/2)$, where $n$ is the center-of-mass excitation and $\nu$ is the relative hyperradial excitation [see Appendix A].

For $N$ equal masses, the Coxeter group is $A_{N-1}$ and the ground state corresponds to the lowest-energy fermionic state in a harmonic trap restricted to a sector (a la Girardeau) as expected~\cite{yukalov_fermi-bose_2005, harshman_spectroscopy_2014}. This equal-mass solution can also be seen as the limiting case of the ground state of the Calogero-Moser model with inverse-square interactions in a harmonic trap with zero coupling constant~\cite{brink_explicit_1992, vacek_eigenfunctions_1994}. However, unlike the equal-mass solutions, the non-equal mass solutions cannot be considered as restrictions of fermionic solutions to a single sector.

In addition to the ground state (\ref{eq:Ups0}), the excited state relative hyperangular wave functions in a Coxeter sector is also constructed using a Bethe ansatz-like superposition of hyperspherical harmonics. Hyperspherical harmonics are homogenous polynomials in $\hat{z}_i$ that are eigenstates of the relative angular momentum $L^2_{rel}$ with eigenvalue $\lambda(\lambda + N - 3)$~\cite{avery_hyperspherical_1989, yanez_position_1994}. The method takes advantage of the fact that reflections $R_{ij}$ commute with $L^2_{rel}$, or in other words the Coxeter group of rank $(N\!-\!1)$ is a subgroup of the orthogonal transformations $\mathrm{O}(N\!-\!1)$ [see Appendix B]. Like (\ref{eq:Ups0}), excited solutions are first constructed over the whole sphere, and then restricted to the Coxeter sectors. By construction, the excited states are antisymmetric with respect to reflections in the Coxeter group. Not all values $\lambda > \lambda_0$ for the relative angular momentum support such solutions. For the three groups $A_3$, $C_3$ and $H_3$, the allowed spectra of $\lambda$ are
\begin{subequations}\label{eq:spec4}
\begin{eqnarray}
&A_3:&\ \lambda = 6 + 3 n_1 + 4 n_2,\\
&C_3:&\ \lambda = 9 + 4 n_1 + 6 n_2,\ \mbox{and}\\
&H_3:&\ \lambda = 15 + 6 n_1 + 10 n_2,\label{eq:spec4:h3}.
\end{eqnarray}
\end{subequations}
In each case, the first number in the sum is $\lambda_0$, the relative angular momentum of the ground state and the number of reflections in the Coxeter group. Then the non-negative integers $n_1$ and $n_2$ label the excited states. Degeneracies  in the hyperangular degrees of freedom arise when multiple pairs of integers provide the same $\lambda$, and the pattern of degeneracies matches the prediction of Weyl's Law for a spherical triangle [see below]. The series of positive integers $3 n_1 + 4 n_2$,  $4 n_1 + 6 n_2$, and $6 n_1 + 10 n_2$ in (\ref{eq:spec4}) correspond precisely to the orders of homogeneous polynomials that have definite relative angular momentum and are \emph{symmetric} under the action of reflections in the groups $A_3$, $C_3$, and $H_3$, respectively~\cite{humphreys_reflection_1992}. Incorporating the center-of-mass and hyperradial degrees of freedom, all energy eigenstates are uniquely identified by four quantum numbers: $\{n, \nu, n_1, n_2\}$.

For $N \geq 4$ and general masses, or for Coxeter masses but in an arbitrary order, we believe the dynamics within sectors are not integrable in any sense. In the case of $H_3$, we provide evidence by numerically solving the spherical Laplacian for Coxeter masses in all sectors. We use the following method~\cite{dehkharghani_impenetrable_2016}: Each spherical triangle is flattened into an isosceles right triangle. The flattening coordinate transformation distorts the spherical Laplacian into a new operator whose spectrum must be solved inside the triangle with hard-wall boundary conditions. The spectrum is found by diagonalizing this transformed Laplacian in a basis of exact solutions for the right triangle. Details about this procedure and its convergence are described in Appendix C.

\begin{figure}
\includegraphics[width= \columnwidth]{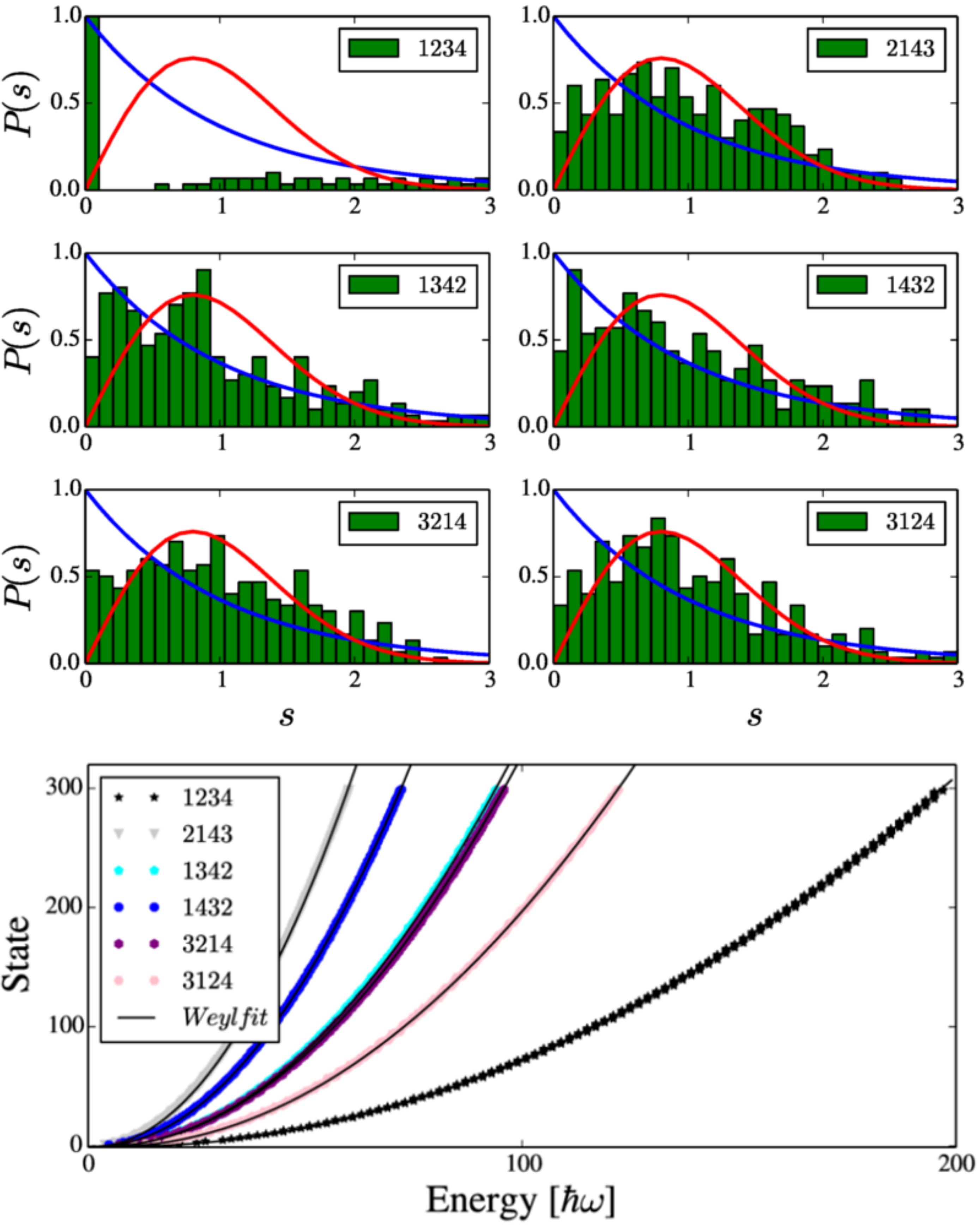}
\caption{This figure depicts unfolded spectrum statistics for $H_3$ Coxeter masses with mass fractions $\mu_1 = \mu_4 =0.44279$, $\mu_2 =  0.03381$, and $\mu_3 = 0.08061$. There are only six different sectors because of two equal masses  $m_1 = m_4$ and because $\Omega_{p_1p_2p_3p_4}$ is congruent to $\Omega_{p_4p_3p_2p_1}$ by inversion. The variable $s$ is the normalized unfolded energy level difference~\cite{bohigas1991}. The integrable sector $\Omega_{1234}$ is depicted in the top left graph and agrees with the prediction from (\ref{eq:spec4:h3}). The blue lines depict the Poissonian statistics expected for an integrable system; the red lines are Wigner-Dyson distribution derived from random matrix theory expected for quantum ergodic systems with time-reversal symmetric Hamiltonians. The bottom graph shows the quality of the Weyl's Law (\ref{eq:weyl}) in the integrable sector $\Omega_{1234}$ as well as the non-integrable sectors.}\label{fig:FFF}
\end{figure}

The level spacing statistics (after
the standard unfolding~\cite{bohigas1991}) for a set of four particles with $H_3$ mass ratios are depicted in Fig.~\ref{fig:FFF}. The first sector depicted is the integrable sector, whose numerical solution agrees with the prediction of (\ref{eq:spec4:h3}), the other five sectors are for the same masses arranged in other orders. For integrable sectors, the unfolded energy level statistics are expected to follow a Poissonian distribution. In our case, within the Coxeter sector the extra degeneracies of the system due to superintegrability distort the distribution so that it is peaked even more strongly at zero-energy gap~\cite{whan_hierarchical_1997}. What we demonstrate in Fig.~\ref{fig:FFF} is that even for Coxeter mass families, the incorrectly-ordered sectors show numerical evidence for quantum ergodicity in the form of Wigner-Dyson distributions for their eigenvalues. 

Our numerical results for the spacing statistics open questions about the transition from integrability to ergodicity. We have performed numerical simulations on a variety of integrable mass families. While the characteristics of the Coxeter sectors remains stable, the other sectors sometimes look closer or further from Wigner-Dyson distributions, as is already visible in Fig.~\ref{fig:FFF}. We have investigated several possibilities for these intermediate distributions, such as integrable subclusters, but we have not arrived at any conclusive results. We have also investigated small random deviations from integrable mass sectors of the order of 5\%. For this scale of deviation, the formerly integrable sectors still look far from Wigner-Dyson, but closer to Poissonian than the energy level statistics for exact Coxeter masses. Understanding the ragged edge between integrability and ergodicity using this model seems to be a productive avenue for future investigation.

To demonstrate that we find all the spectrum from this procedure, we have compared our results for $N(\tilde{E})$, the total number of energy eigenvalues below scaled energy $\tilde{E}$, to the prediction of Weyl's Law~\cite{ivrii_100_2016} for a sphere:
\begin{equation}\label{eq:weyl}
N(\tilde{E}) = \frac{A}{4\pi} \tilde{E} - \frac{\ell}{4 \pi} \sqrt{\tilde{E}},
\end{equation}
where $\tilde{E} = (2mR^2/\hbar^2) E$ and $R$ is an arbitrary radius. The second term is the correction due to the Dirichlet boundary conditions proportional to the boundary length $\ell$. The area of spherical triangle $\Omega_{ijkl}$ is $A/R^2= \omega_{ijk} + \omega_{jkl} +\omega_{ij,kl} - \pi$ (Girard's Theorem~\cite{brooks_spherical_2005}). The perimeter is $\ell/R = \varphi_{ijk} + \varphi_{jkl} + \varphi_{ij,kl}$, where the vertex angles ($\varphi_{ijk}$, $\varphi_{jkl}$, $\varphi_{ij,kl}$) satisfy~\cite{weisstein_spherical_nodate}
\begin{equation}
\cos \varphi_{ijk} = \frac{\cos \omega_{ijk} + \cos \omega_{jkl} \cos \omega_{ij,kl}}{\sin \omega_{jkl} \sin \omega_{ij,kl}}
\end{equation}
and cyclic permutations of ($\varphi_{ijk}$, $\varphi_{jkl}$, $\varphi_{ij,kl}$) and ($\omega_{ijk}$, $\omega_{jkl}$, $\omega_{ij,kl}$). Fig.~\ref{fig:FFF} compares numerical solutions to this prediction.

\section*{Liouville integrability and superintegrability}

The  separability of the model (\ref{eq:hamx}) provides four functionally-independent integrals of the motion for $N\geq 3$ particles with any masses: the center-of-mass Hamiltonian $H_{\COM}$, the relative Hamiltonian $H_{\rel}$, the total angular momentum squared $L^2$, and the relative angular momentum squared $L_{\rel}^2$. Three of these integrals $\{H_{\COM}, H_{\rel}, L_{\rel}^2\}$ are in involution, but $L^2$ does not commute with them. This set of four integrals of motion is sufficient to prove integrability and superintegrability (but not maximal superintegrability) for $N=3$ with any masses. For $N=4$ this is not enough to prove integrability, which requires four integrals in involution, nor is it enough for superintegrability, requiring at least five total conserved quantities, and certainly not enough for maximal superintegrability, requiring seven conserved quantities.

Nonetheless, we conjecture that our model is maximally superintegrable for $N \geq 3$ when in a sector with Coxeter mass ratios. Our evidence is: 
(1) using the method of images~\cite{hobson_ergodic_1975, richens_pseudointegrable_1981}, the classical problem can be shown to support closed orbits indicating maximal superintegrability; (2) the correspondence of our model as a limiting case of certain CSM models which are maximally superintegrable~\cite{wojciechowski_superintegrability_1983, hakobyan_superintegrability_2014}. The general construction of a set of observables that provide maximal superintegrability for 
all reflection groups in any number of spatial dimensions requires the methods of algebraic geometry. This ongoing project will be the subject of a future publication.

Let us outline a potential strategy for searching for the missing integrals of motion for four particles using $H_3$ as an example. The key role is played by the invariant polynomials of the group  $q_m(z_{1}, z_{2}, z_{3})$ with order $m$. These are the lowest-order, homogeneous, functionally-independent polynomials that remain unchanged  under any of the group transformations. They are known and tabulated for all the reflection groups~\cite{mehta_basic_1988}. For $\mbox{H}_3$,  there are three invariant polynomials $q_m$ with order $m=2$, $6$ and $10$ and (up to a normalization) they are constructed as
\begin{equation}
q_{m}(z_{1}, z_{2}, z_{3}) = \sum_{\{  \bm{\sigma}\}} 
( \bm{\sigma}\cdot(z_{1},\,z_{2},\,z_{3}))^{m},
\end{equation}
where $\{\bm{\sigma}\}$ are the set of vectors describing the six five-fold rotation axes of $H_3$. From the three polynomials $q_2$, $q_6$, and $q_{10}$, we define the three operators $J_m \equiv q_{m}(L_{12},\,L_{23},\,L_{31})$. Here, $L_{ij}$ are components of the vector of the relative  angular momentum in the $ij$-plane. Note that $J_{2}$ is proportional to $L_{\rel}^2$ and so it does not give an additional integral of motion.

However, the operator $J_{6}$ completes the commuting set $\{H_{\COM}, H_{\rel}, L_{\rel}^2 \}$ to a Liouvillian set.  Since $J_{6}$ commutes with mirror reflections of the $H_3$ group, by Schur's Lemma it must act as a multiple of the identity on the antisymmetric states. It commutes with the previous three members of the Liouvillian set, and all four can be readily shown to be functionally-independent in the classical sense. The five-member set
$\{H_{\COM}, L^2, H_{\rel}, L_{\rel}^2, J_{6} \}$ now establishes superintegrability for the $H_3$ mass family.

This set can be further extended to a maximally superintegrable set using the operator $J_{10}$ and another invariant operator $I_{6}$ defined by
\begin{equation}
I_{6} \equiv q_{6}(a^{\dagger}_{1},\,a^{\dagger}_{2},\,a^{\dagger}_{3}) q_{6}(a_{1},\,a_{2},\,a_{3}),
\end{equation}
where 
$a_{j} = (-i\partial_{z_{j}} - i z_{j})/\sqrt{2}$ is an annihilation operator for the $j$-th component of the relative motion. The 
operator $I_6$ naturally commutes with the total Hamiltonian $H$. The resulting seven-member set is (classically) functionally independent and establishes maximal superintegrability for the $H_3$ model. 
The scheme can readily be generalized to the other two three-dimensional reflection groups, $A_3$ and $C_3$. However, no ready generalization to higher dimensions exists for the Liouvillian sets, because, a priori, the operators $J_{m}$ do not commute between themselves. Finding Liouvillian sets for higher-dimensional groups is a subject of future work. Identifying and classifying the maximal superintegrablty sets, and ideally connecting them to the
known integrals for the Calogero-Moser model~\cite{wojciechowski_superintegrability_1983, hakobyan_superintegrability_2014, saghatelian_constants_2012}, is another ongoing project. 

\section*{Experimental outlook}

At the moment, three possible experimental applications of the models considered in this article can be foreseen. The first possibility is the straightforward idea of finding a collection of atoms that naturally have the right mass ratios and seeking the signatures of integrability in the spectral, coherence, and thermalization properties of the system. Even if the particles are only close to Coxeter family, our numerical results for the energy spectrum suggest that traces of integrability should still be present. More generally, the Coxeter criteria can be used to measure how far from integrability particular arrangements of imbalanced masses are expected to be, or whether there are integrable subclusters possible within a multi-species ultracold atomic gas.

In the second scheme, if real masses with the correct ratio are not available, the atomic mass is controlled using optical lattices. Given sufficient laser power, the effective mass \cite{kittel_introduction_2004} can be tuned from its ``bare'' value to almost zero \cite{gadway_glassy_2011}. In particular the effective mass can be made $3$ times greater than its bare counterpart in a lattice of a depth $V_{0} = 7 E_{\text{R}}$, and $23$ times greater for $V_{0} = 16 E_{\text{R}}$
respectively. Here $E_{\text{R}} = \hbar^2 k^2/(2 m)$ is the so-called recoil energy, $k$ is the wave vector of light that creates the lattice, $m$ is the bare atomic mass. In both cases, harmonic confinement represents the most natural experimental
environment, unlike the box and ring geometries traditionally studied using Bethe ansatz methods. 

In the third scheme, described in more detail in \cite{olshanii_creating_2016}, the role of massive particles is played by bosonic solitons in an atom waveguides \cite{khaykovich_formation_2002, strecker_formation_2002, cornish_formation_2006}. The solitons are made of atoms in two alternating internal states where the intraspecies interaction is attractive and the interspecies interaction is repulsive. The goal would be to engineer clusters of atoms whose combined masses satisfy the Coxeter criteria. For example, the clustering pattern $(3m,m,2m,6m)$ has $C_3$ symmetry. A mixture of ${}^{7}\mbox{Li}$ atoms with $m_{F}=-1$ and $m_{F}=0$  in a magnetic field of $855\, \mbox{G}$ constitutes an example~\footnote{R.G.~Hulet, private communication.}.

Any implementation of the models considered in our article may constitute an efficient experimental realization of spherical triangular (or higher-dimensional, simplex-shaped) quantum billiards~\footnote{We note that in principle the relative Hamiltonian for any of the $N\!=\!4$ families also could be realized with a non-interacting Bose-Einstein condensate in a spherically-symmetric harmonic trap sliced by six sheets of intense laser light, or even just three sheets to create a sector.}. The ergodicity of classical flat triangular billiards is conjectured to strongly depend on the rationality of the billiard angles~\cite{casati_mixing_1999, wang_nonergodicity_2014}.  Numerically, such questions about ergodicity are difficult, requiring long propagation for averages to converge to their infinite time limits. A study of the eigenstate-to-eigenstate variance of the expectation values of observables~\cite{deutsch_quantum_1991, srednicki_chaos_1994, rigol_thermalization_2008}, which is a faithful quantum analogue of classical deviations from ergodicity (c.f.~\cite{zhang_atom_2015} for a comparison), may provide an efficient alternative to classical long-time averages. Experimentally, one may conjecture an appearance of a memory of initial conditions, if the billiard is not ergodic~\cite{yurovsky_memory_2011}. The mass mixtures considered in our paper could constitute a way to study multi-dimensional classical and quantum hard-wall billiards with continuously tunable geometry, a powerful extension of the existing experimental techniques~\cite{friedman_observation_2001}.

\emph{Acknowledgments}: This work was supported by the Aarhus University Research Foundation, the US National Science Foundation Grant No. PHY-1607221, the Office of Naval Research Grant N00014-12-1-0400, The Danish Council for Independent Research (Sapere Aude), and the Humboldt Foundation. Thanks to FF Bellotti, MES Andersen, L Rammelm\"uller, and F Werner for useful conversations.

\section*{Appendix A: Coordinate Transformations and the Map to Quantum Billiards}

In the following section we establish the map from the model Hamiltonian (\ref{eq:hamx})
to a free particle in a bounded region on the $(N\!-\!2)$-sphere. Much of this is well-known~\cite{werner_unitary_2006, dehkharghani_impenetrable_2016}, but we reproduce it here for the readers' convenience and to establish notation.

\begin{figure}
\adjustbox{trim={.2\width} {.41\height} {0.25\width} {.35\height},clip}{\includegraphics[width= 1.8\columnwidth]{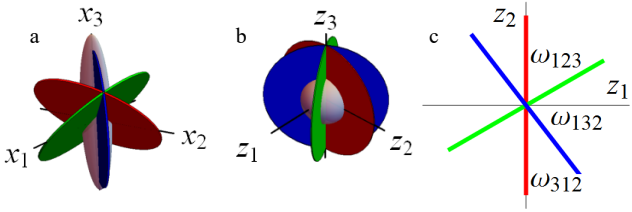}}
\caption{This figure depicts the coordinate transformation for three particles with the $A_2$ masses $m_1= 1/13 M$, $m_2 =9/52 M$, and $m= 3/4 M$, where $M$ is the total mass. Subfigure (a) depicts the configuration space in natural particle positions ${\bf x}$; the gray ellipsoid represents an equipotential for the equal-frequency (and therefore not equal-strength); the red, green, and blue disks represent the $X_{12}$, $X_{23}$, and $X_{31}$ coincidence planes, respectively. Subfigure (b) depicts the configuration space in mass-scaled and rotated coordinates ${\bf z} = J {\bf x}$, with a spherical equipotential and transformed coincidence planes $Z_{12}$, $Z_{23}$, and $Z_{31}$. Subfigure (c) depicts the project of (b) into relative coordinate $z_1$-$z_2$ plane. The sector with angle $\omega_{123} = \pi/3$ is the Coxeter sector.}\label{fig:ggg}
\end{figure}

The equipotentials of (\ref{eq:hamx}) are $N$-dimensional ellipsoids segmented into $N!$ sectors by $(N-1)$-dimensional hyperplanes $X_{ij}$ defined by the particle coincidences $x_i -x_j=0$; see Fig.\ \ref{fig:ggg}. In the limit $g \to \infty$ these planes are impenetrable. The angle between coincidence hyperplanes $X_{ij}$ and $X_{jk}$ that share a particle is $\pi/3$ (only possible for $N\geq 3$); the angle between hyperplanes $X_{ij}$ and $X_{kl}$ that do no share a particle is $\pi/2$ (only possible for $N \geq 4$).

As a first step, we scale the position coordinates into unitless position variables $y_i = \sqrt{m_i \omega /\hbar} \, x_i$. Then the Hamiltonian (\ref{eq:hamx}) becomes
\begin{eqnarray}\label{eq:hamy}
H &=& \frac{\hbar\omega}{2} \sum_{i=1}^N \left( - \frac{\partial^2}{\partial y_i^2} + y_i^2 \right) \\
&& +   \sum_{i<j} \tilde{g}_{ij} \delta\left(  \sqrt{\frac{\mu_{ij}}{m_i}} y_i - \sqrt{\frac{\mu_{ij}}{m_j}} y_j \right),\nonumber
\end{eqnarray}
where $\mu_{ij} = m_i m_j/(m_i + m_j)$ and $\tilde{g}_{ij} = g_{ij} \sqrt{\mu_{ij}\omega/\hbar}$. This scaling transformation ${\bf y} = S {\bf x}$ has brought the harmonic potential into a form with $N$-spherical symmetry but at the cost of de-symmetrizing the coincidence planes. To describe the geometry, define the transformed coincidence planes $Y_{ij} \equiv S X_{ij}$ with normals $\hat{\bbeta}_{ij}$. The contact interaction then has the form 
\begin{equation}\label{eq:inty}
\sum_{i<j} \tilde{g}_{ij} \delta\left( \hat{\bbeta}_{ij}\cdot {\bf y}  \right).
\end{equation}
The angle $\omega_{ijk}$ between coincidence planes $Y_{ij}$ and $Y_{jk}$ with normals $\hat{\bbeta}_{ij}$ and $\hat{\bbeta}_{jk}$ is now
\begin{equation}
\omega_{ijk} = \arctan\left( \sqrt{\frac{m_j(m_i+m_j+m_k)}{m_i m_k}} \right).
\end{equation}
Whereas in the equal mass case, $\omega_{ijk}$ is always $\pi/3$, for three arbitrary masses it can range from $0$ ($m_j$ much lighter than other two masses) to $\pi/2$ ($m_j$ much heavier). The angle $\omega_{ij,kl}$ between coincidence planes $Y_{ij}$ and $Y_{kl}$ that do not share a particle remains $\pi/2$.  

In the limit $g \to \infty$ the Hamiltonian (\ref{eq:hamy}) separates in hyperspherical coordinates with radius $R^2 = \sum_i y_i^2$. The interaction term [\ref{eq:inty}] is proportional to $1/R$, so it is not separable for finite values of $\tilde{g}_{ij}$, but as $\tilde{g}_{ij} \to \infty$ there is no distinction between $1/R$ or $1/R^2$ times the sum of delta functions. So hyperspherical symmetry of (\ref{eq:hamy}) emerges and there is $\mathrm{SO}(2,1)$ dynamical symmetry in the total hyperradial coordinate $R$~\cite{werner_unitary_2006}. To develop physical intuition, it is sometimes useful to imagine a single, classical particle bouncing around in this $N$-dimensional landscape. In this mass-rationalized geometry, the classical particle trajectory changes its direction of angular momentum when it bounces off of a coincidence hyperplane $Y_{ij}$, but it does not change its magnitude of ``angular momentum'' in configuration space. However, total angular momentum does not respect the center-of-mass separability and does not commute with the relative Hamiltonian or relative angular momentum [see below], so we do not exploit it here.
 
Next we rotate the coordinate system ${\bf z} = J {\bf y}$ so that the component $z_N\equiv Z$ is the scaled center-of-mass $Z = \sum_i  y_i \sqrt{m_i/M}$ and $M$ is the total mass. The orthogonal transformation $J$ with this property is not unique and its selection determines a particular choice for Jacobi relative coordinates $z_1$ through $z_{N-1}$. The transformation $J$ also rotates the coincidence planes $Z_{ij} \equiv J Y_{ij}$  and their normals $\hat{\bgamma}_{ij} \equiv J \hat{\bbeta}_{ij}$, but leaves the angles between planes like $\omega_{ijk}$ and $\omega_{ij,kl}$ invariant. Since all the normal vectors $\hat{\bgamma}_{ij}$ have zero $Z$-components, the Hamiltonian in ${\bf z}$-coordinates separates into $H = H_{\COM} + H_{\rel}$, where $H_{\COM}$ is the Hamiltonian for a one-dimensional harmonic oscillator in the center-of-mass $Z$ coordinate and the relative Hamiltonian is
\begin{equation}\label{eq:hamz}
H_{\rel} = \frac{\hbar\omega}{2} \sum_{i=1}^{N-1} \left( - \frac{\partial^2}{\partial z_i^2} + z_i^2 \right) +   \sum_{i<j} \tilde{g}_{ij} \delta\left( \hat{\bgamma}_{ij} \cdot {\bf z} \right).
\end{equation}

Finally, we go to hyperspherical coordinates in the relative space, where the relative hyperradius $\rho$ is 
\begin{equation}
\rho^2 = \sum_{i=1}^{N-1} z_i^2 =\sum_{i=1}^N y_i^2   - Z^2,
\end{equation}
and there are $(N\!-\!2)$ angles charting the sphere $\mathcal{S}^{N\!-\!2}$, conventionally chosen as $\Omega = \{\phi, \theta_1, \ldots, \theta_{N\!-\!3}\}$ with $\phi\in [0, 2\pi)$ and $\theta_i \in [0,\pi]$. The relative Hamiltonian now becomes
\begin{eqnarray}\label{eq:hamrho}
H_\rel &=& \frac{\hbar\omega}{2} \left( -\frac{1}{\rho^{N-2}} \frac{\partial}{\partial \rho}\left(\rho^{N-2} \frac{\partial}{\partial \rho} \right) - \frac{1}{\rho^2} \Delta_\Omega + \rho^2 \right)\nonumber\\
&&{} +   \sum_{i<j} \frac{\tilde{g}_{ij}}{\rho} \delta\left( \hat{\bgamma}_{ij} \cdot \hat{z} ,\right)
\end{eqnarray}
where $\Delta_{\Omega}$ is the angular part of the Laplacian in relative configuration space. As before, the relative Hamiltonian (\ref{eq:hamrho}) is $\rho$-$\Omega$ separable in the limit $\tilde{g}_{ij} \to \infty$. 
A general energy eigenstate can be separated into a product of center-of-mass $\zeta(Z)$, relative hyperradial $R(\rho)$ and relative hyperangular $\Upsilon(\Omega)$ functions
\begin{equation}
\Phi_{n,\nu,\lambda,\mu}(Z,\rho,\Omega) = \zeta_n(Z) R_{\nu \lambda} (\rho) \Upsilon_{\lambda,\mu}(\Omega),
\end{equation}
where $n$ is the center-of-mass quantum number, the function $\zeta_n(Z)$ is the one-dimensional harmonic oscillator wave function, and $\nu$ is the relative hyperradial quantum number. At this point, $\lambda$ is just derived from the judiciously-parametrized  relative hyperangular separation constant $\lambda(\lambda + N - 3)$ and $\mu$ is just an additional label to distinguish any possible degenerate states for a given $\lambda$. 
If there was no angular potential $\tilde{g}_{ij}=0$, then $\lambda$ would be a non-negative integer, the eigenfunctions on $\mathcal{S}^{N\!-\!2}$ would be the hyperspherical harmonics, and $\mu$ would be a collective index to label degeneracies~\cite{avery_hyperspherical_1989, yanez_position_1994}. However since there is an angular potential in (\ref{eq:hamrho}), the hyperangular solutions are unknown and we must explicitly solve for $\lambda$, including any possible degeneracies. Whatever value $\lambda$ takes (including non-integer values), the relative hyperradial function $R_{\nu \lambda}(\rho)$ is the standard solution for the radial factor of an $(N\!-\!1)$-dimensional isotropic harmonic oscillator in hyperspherical coordinates~\cite{yanez_position_1994} with energy $\hbar\omega(2 \nu + \lambda + (N\!-\!1)/2)$:
\begin{equation}
R_{\nu \lambda}(\rho) = A_{\nu,\lambda} \rho^\lambda L^{\lambda + \frac{N\!-\!3}{2}}_\nu(\rho^2) e^{-\rho^2/2},
\end{equation}
 where $A_{\nu,\lambda} = \sqrt{2\,\nu!/\Gamma(\nu\! +\! \lambda \!+\! (N\!-\!1)/2)}$.

We have achieved our desired result: this series of coordinate transformations has reduced solving the $N$-particle Hamiltonian (\ref{eq:hamx}) with equal frequencies and infinite-strength contact interactions into solving hard-wall quantum billiards in $(N\!-\!2)$-simplexes on $(N\!-\!2)$-spheres. 

\section*{Appendix B: Construction of Exact Solutions}

In this section, we construct the wave functions within Coxeter sectors, i.e.\ sectors of the $\mathcal{S}^{N\!-\!2}$ hypersphere defined by the relative hyperangular coordinates that have the right shape to tile the sphere under reflection. For convenience, choose the Coxeter sector to be the ordering sector $\Omega_{12\cdots N}$ so that it is bounded by the coincidence hyperplanes $Z_{12}$ through $Z_{(N\!-\!1)N}$. These $(N\!-\!1)$ hyperplanes define the reflections $R_{12}$ though $R_{(N\!-\!1)N}$ that generate the Coxeter group.

The Coxeter group $G_m$ is generated by $m$ reflections in $m$ dimensions. As such, it can considered as a subgroup of $\mathrm{O}(m)$, orthogonal transformations in $m$ dimensions and the symmetry of the sphere $\mathcal{S}^{m\!-\!1}$. To summarize the method, there is a solution to the Hamiltonian in the Coxeter sectors $\Omega_{12\cdots N}$ and $\Omega_{N\cdots 21}$ whenever an irreducible representation (irrep) of $G_{N\!-\!1}$ that is antisymmetric under all reflections appears in the decomposition of an irrep of $\mathrm{O}(N\!-\!1)$. The irreps of $\mathrm{O}(m)$ generally are reducible with respect to the subgroup $G_m$. The method of characters can answer the question as to whether an irrep of a subgroup appears in the decomposition of the irrep of the group. When it does exist, the corresponding states can be constructed using projection operators and (in the case of degeneracies) an orthonormalization procedure.

The irreducible representations for $\mathrm{O}(m)$ and their realizations by hyperspherical harmonics are well-known~\cite{hamermesh_group_1989, wybourne_classical_1974}, and so we just summarize a few facts here for the readers' convenience. The subgroup $\mathrm{SO}(m)$ is a Lie group with $m(m-1)/2$ generators in the Lie algebra. Denote these generators as $L_{ij}$ for $i< j$ with $i,j\in\{1,\ldots,m\}$, where $L_{ij}$ generates a rotation in the $ij$-plane. The quadratic Casimir of $\mathrm{SO}(m)$ is the sum of all of these generators squared
\[
L^2 = \sum_{\langle i,j \rangle} L_{ij}^2.
\]
For $\mathrm{SO}(3)$ this is the familiar angular momentum squared operator with eigenvalues $\lambda(\lambda +1)$.  The $\mathrm{SO}(3)$ irreps are labeled by $\lambda$ and have degeneracy $2\lambda+1$. In $m > 3$ dimensions, the operator $L^2$ is the hyperangular momentum squared operator with eigenvalue $\lambda(\lambda + m - 2)$ and an irrep labeled by $\lambda$ has degeneracy~\cite{avery_hyperspherical_1989}
\[
d(\lambda) = \frac{(m+2\lambda -2)(m+\lambda-3)!}{\lambda!(m-2)!}.
\]
Once the representation of total inversion is chosen, the irreps of $\mathrm{SO}(m)$ also naturally carry a representation of $\mathrm{O}(m)$ . For example,  inversion is represented in the $\lambda$ irrep by multiplication by $(-1)^\lambda$ for $\mathrm{O}(3)$.

To reduce an $\mathrm{O}(m)$ irrep $\lambda$ into the irreps of Coxeter group $G_m$, the $G$ elements are sorted into conjugacy classes $K_i$ with $k_i$ elements.  Each irrep $W$ of $G_m$ has a unique pattern of characters $\chi^{W}(K_i)$. In particular we are interested in the $G_m$ irrep $W=A$ of all anti-invariant states, meaning $\chi^A(K_i) = 1$ when $K_i$ is a conjugacy class whose elements are an even composition of reflections and $\chi^A(K_i) = -1$ when $K_i$ is a class comprised of odd compositions. Further, each conjugacy class has a character $\chi^\lambda(K_i)$ in the $G_m$-reducible $\mathrm{O}(m)$-irrep denoted by $\lambda$.  When these characters are known, then the number of times the $G_m$ irrep $A$ appears in the decomposition of the $\mathrm{O}(m)$ irrep $\lambda$ is~\cite{hamermesh_group_1989}
\begin{equation}\label{eq:alambda}
a_\lambda = \frac{1}{G} \sum_{K_i} k_i \chi^A(K_i) \chi^{\lambda}(K_i).
\end{equation}
Note that $\chi^A(K_i)=(\chi^A(K_i))^*$ because Coxeter groups are ambivalent.
The number $a_\lambda$ is an integer that counts how many solutions there are with relative angular momentum $\lambda$. The projection operator onto the anti-invariant irrep $A$ is given by
\begin{equation}\label{eq:proj}
P^A = \frac{1}{G} \sum_{g\in G_m} \chi^A(g) D^\lambda(g),
\end{equation}
where $D^\lambda(g)$ is the representation of group element $g$ acting irreducibly on the $d(\lambda)$-dimensional representation space. If $a^\lambda=1$, any vector in the $\lambda$ irrep space with a non-zero projection will be proportional to the solution we seek.   If $a_\lambda>1$, then a set of orthonormal solutions can be found by projecting multiple vectors and then applying Gram-Schmidt orthogonalization.

\begin{table}
\caption{This table provides information about the conjugacy classes of $H_3$. The first column is the Sch\"onflies notation for the elements in the  class $K_i$. The second column gives the angle of rotation $\phi_i$ of the element realized in $\mathrm{O}(3)$. The third column is whether it is generated by an even or odd number of reflections $\pi_i = \pm 1$. All odd elements that are rotations can be considered as rotoreflections, i.e.~a rotation followed by a reflection in the plane perpendicular to the rotation axis. The fourth and fifth columns are the order of the elements in the class $K_i$ and the number of elements $k_i$ in the class.}
\centering
\label{tab:H3}
\begin{tabular}{|c|c|c|c|c|}
\hline
Elements  & Angle & Even/odd & Order  & Number\\
\hline
$E$ & $0$ & $+$ & 1 & 1\\
$C_5,C_5^4$ & $2\pi/5$ & $+$ & 5 & 12\\
$C_5^2,C_5^3$ & $4\pi/5$ & $+$  & 5 & 12\\
$C_3,C_3^2$ & $2\pi/3$ & $+$  & 3 & 20\\
$C_2$ & $\pi$ & $+$  & 2 & 15\\
$I$ & $\pi$ & $-$  & 2 & 1\\
$S_{10},S_{10}^9$ & $\pi/5$ & $-$  & 10 & 12\\
$S_{10}^3,S_{10}^7$ & $3\pi/5$ & $-$  & 10 & 12\\
$S_6,S_6^5$ & $\pi/3$ & $-$  & 6 & 20\\
$\sigma$ & $0$ & $-$  & 2 & 15\\
\hline
\end{tabular}
\end{table}

As an example, consider the Coxeter group $H_3$. This group has ten conjugacy classes summarized in Table~\ref{tab:H3}. The $H_3$ character $\chi^A(K_i)$ for the anti-invariant irrep is $+1$ for the five even conjugacy classes and $-1$ for the five odd classes. The $\mathrm{O}(m)$ character $\chi^\lambda(K_i)$ for irrep $\lambda$ is 
\begin{equation}
\chi^\lambda(K_i) = \sum_{\mu=-\lambda}^\lambda \cos(m \phi_i) (\pi_i)^{\lambda - \mu},
\end{equation}
where $\phi_i$ is the angle of rotation and $\pi_i$ is the reflection parity for the conjugacy class $K_i$. Plugging this into (\ref{eq:alambda}), we find the pattern of degeneracies given in eq.~[6c] of the main text. The same method is used in \cite{harmer_spectra_2007} to find which $\mathrm{O}(m)$ irreps have symmetric irreps of the spherical triangle groups in their reduction. 

The construct the actual states $\Upsilon_\lambda(\theta,\phi)$, we use the projection operator (\ref{eq:proj}) acting on the spherical harmonics. Instead of explicitly constructing the $(2\lambda +1) \times (2\lambda +1)$ unitary matrices $D^\lambda(g)$ that act on the spherical harmonics $Y^\mu_\lambda(\theta,\phi)$ for each of the 120 elements of $H_3$, we choose a slightly different method that takes advantage of two facts: (1) the spherical harmonics can be written as polynomials of the relative coordinates; and (2) we already have the $3\times 3$ matrices $O(g)\in\mathrm{O}(3)$ that represent $H_3$ as rotation and reflections.

The first step is to express the spherical harmonics in terms of the relative coordinates $(z_1,z_2, z_3)$. We work with the real form of the spherical harmonics, defined as
\begin{eqnarray*}
 Y_{\lambda\mu}(\theta,\phi) = \left\{ \begin{array}{ll} \sqrt{2} N_{\lambda\mu} P^\mu_\lambda(\cos\theta)\cos(\mu\phi) & \mu > 0\\
N_{\lambda 0} P^0_\lambda(\cos\theta) & \mu = 0\\
\sqrt{2} N_{\lambda\mu} P^{|\mu|}_\lambda(\cos\theta)\sin(|\mu|\phi) & \mu < 0 \end{array} \right. 
\end{eqnarray*}
where
\[
 N_{\lambda\mu} = \sqrt{\frac{2 \lambda +1}{4 \pi} \frac{(\lambda - |\mu|)!}{(\lambda + |\mu|)!}}.
\]
The real spherical harmonics can be written in terms of the relative coordinates:
\[
Y_{\lambda\mu}(\hat{\bf z}) \equiv Y_{\lambda\mu}(\cos^{-1}(z_3/\rho),\tan^{-1}(z_2/z_1)).
\]
 Noting the relations
\begin{eqnarray*}
\cos(\mu\phi) &=& (z_1^2+z_2^2)^{-\frac{\mu}{2}} \sum_{k=0}^\mu \left( \!\!\begin{array}{c} \mu \\ k \end{array} \!\!\right) z_1^{\mu-k} z_2^k \cos\frac{\pi k}{2}, \\
\sin(|\mu|\phi) &=& (z_1^2+z_2^2)^{-\frac{\mu}{2}} \sum_{k=0}^\mu \left(\!\! \begin{array}{c} \mu \\ k \end{array}\!\! \right) z_1^{\mu-k} z_2^k \sin\frac{\pi k}{2},\ \mbox{and}\\
P^\mu_\lambda(x) &=& (-1)^\mu (z_1^2+z_2^2)^{\frac{\mu}{2}}  \frac{d^\mu}{d(z_3/\rho)^\mu} P_\lambda(z_3/\rho),
\end{eqnarray*}
we can show that $\rho^\lambda Y_{\lambda\mu}(\hat{\bf z})$ are homogeneous polynomials of order $\lambda$ in $(z_1,z_2, z_3)$.

The projection (\ref{eq:proj}) is applied using the transformation matrix $O(g)$ 
\begin{equation}\label{eq:proj2}
 \rho^\lambda P^A Y_{\lambda\mu}(\hat{\bf z}) =\frac{ \rho^\lambda}{G} \sum_{g\in G_m} \chi^A(g) Y_{\lambda\mu}(O(g)\hat{\bf z}).
\end{equation}
This projection will be zero unless $\lambda$ is in the spectrum given by eq.~[6c] in the main text. Note that it may also be zero for any particular $\mu$, but there must be as many linearly-independent polynomials (that also solve the spherical Laplacian) as there are solutions for $n_1$ and $n_2$ for a given $\lambda$ in eq.~[6c]. For $H_3$, the first time there are multiple solutions is when $\lambda=45$. Explicit calculation for $\lambda=\lambda_0 = 15$ confirms eq.~[5] from the main text for the ground state of $H_3$, and we have performed the same calculations for the other $N=4$ Coxeter groups.


\section*{Appendix C: Numerical Method for Solving Spherical Triangle}

Here we present the numerical method to calculate the energy spectrum for $N=4$ particles with arbitrary masses, which is an extension of the method found in \cite{dehkharghani_impenetrable_2016} to spherical triangles. The general strategy was outlined in the main text: here we provide details about the coordinates, the flattening, and the exact diagonalization used to construct Fig.~4.

After separation of variables, we must solve for the hyperangular wave function $\Upsilon(\Omega)$ within a sector $\Omega_p$. This function must satisfy
\begin{equation}\label{eq:upslap}
\Delta_\Omega \Upsilon(\Omega) = \lambda(\lambda+1) \Upsilon(\Omega)
\end{equation}
with Dirichlet boundary conditions on the three bounding coincidence planes $Z_{p_1p_2}$, $Z_{p_2p_3}$, and $Z_{p_3p_4}$. To solve the problem for arbitrary masses, a particular rotation $J$ in ${\bf z} = JS{\bf x}$ must be specified. For numerical simplicity, we choose the $H$-type four-body coordinates~\cite{rittenhouse_hyperspherical_2011}. Then the rotation $J$ aligns the coordinate plane $Z_{12}$  with the plane $z_1=0$ and aligns the coordinate plane  $Z_{34}$ with $z_2=0$ (see Fig.~1 of main text). The other four coincidence planes are given by the following equations:
\begin{eqnarray*}
Z_{13}:\ \sqrt{\frac{{\scriptstyle m_2(m_3+m_4)}}{{\scriptstyle m_1M}}}z_1 -\sqrt{\frac{{\scriptstyle m_4(m_1+m_2)}}{{\scriptstyle m_3M}}}z_2 + z_3 &=& 0\\
Z_{14}:\ \sqrt{\frac{{\scriptstyle m_2(m_3+m_4)}}{{\scriptstyle m_1M}}}z_1 + \sqrt{\frac{{\scriptstyle m_3(m_1+m_2)}}{{\scriptstyle m_4M}}}z_2 + z_3 &=& 0\\
Z_{23}:\ \sqrt{\frac{{\scriptstyle m_1(m_3+m_4)}}{{\scriptstyle m_2M}}}z_1 + \sqrt{\frac{{\scriptstyle m_4(m_1+m_2)}}{{\scriptstyle m_3M}}}z_2 -  z_3 &=& 0\\
Z_{24}:\ \sqrt{\frac{{\scriptstyle m_1(m_3+m_4)}}{{\scriptstyle m_2M}}}z_1 - \sqrt{\frac{{\scriptstyle m_3(m_1+m_2)}}{{\scriptstyle m_4M}}}z_2 -  z_3 &=& 0.
\end{eqnarray*}

The numerical problem is solved independently in each sector $\Omega_{p_1p_2p_3p_4}$ and checked for consistency with the similar sector $\Omega_{p_4p_3p_2p_1}$. In the following, as an example we solve the sector $\Omega_{1342}$, which is limited by $Z_{34}$, $Z_{13}$, and $Z_{24}$ coincidence planes. We isolate $z_1$ in the above equations and introduce (non-standard) spherical coordinates with $z_1=\rho\cos\theta$, $z_2=\rho\sin\theta \cos\phi$ and $z_3=\rho\sin\theta \sin\phi$ so that $\cos\theta > 0$ in sector $\Omega_{1342}$. Then substituting spherical coordinates into the coincidence plane equations and dividing by $z_1 = \rho\cos\theta > 0$, we find
\begin{eqnarray}
\tan\theta\cos\phi &=& 0\nonumber\\
1-a\tan\theta\cos\phi+b\tan\theta\sin\phi &=& 0\nonumber\\
1-c\tan\theta\cos\phi-d\tan\theta\sin\phi &=& 0
\end{eqnarray}
with
\begin{eqnarray*}
&&a=\sqrt{\frac{(m_1+m_2) m_1 m_4}{(m_3+m_4) m_3 m_2}},\ \ b=\sqrt{\frac{M m_1}{(m_3+m_4) m_2}},\\
 && c=\sqrt{\frac{(m_1+m_2) m_2 m_3}{(m_3+m_4) m_1 m_4}},\ \mbox{and} \ 
d=\sqrt{\frac{M m_2}{(m_3+m_4) m_1}}. 
\end{eqnarray*}

Introducing new coordinates $u=\tan\theta\cos\phi$ and $v=\tan\theta\sin\phi$, the boundary conditions 
\begin{eqnarray*}
u&=&0\\
1-a u+bv &=&0 \\
1-c u-dv &=&0.
\end{eqnarray*}
are simple and  describe a triangle in flat space. However, the differential operator $\Delta_{\Omega}$ has become more complicated:
\begin{eqnarray}\label{eq:deltauv}
\Delta_{\Omega} &=& (1+u^2+v^2)\left\{ (1+u^2)\frac{\partial^2}{\partial u^2}+(1+v^2)\frac{\partial^2}{\partial v^2} \right.\nonumber\\
&& \left. {} + (2uv)\frac{\partial^2}{\partial u\partial v}+(2u)\frac{\partial}{\partial u}+(2v)\frac{\partial}{\partial v} \right\}.
\end{eqnarray}

Next we introduce a final coordinate transformation 
\begin{eqnarray*}
s &=& \frac{2d}{(b+d)}(-au+bv)-\frac{(b-d)}{(b+d)}\\
t &=& \frac{2b}{(b+d)}(-cu-dv)-\frac{(d-b)}{(b+d)}.
\end{eqnarray*}
This can be inverted as
\begin{eqnarray*}
u &=& -\frac{(b+d)}{2(ad+bc)}(s + t)\\
v &=& -\frac{a(b+d)}{2b(ad+bc)}(s+t)+\frac{(b-d)+(b+d)s}{2bd}.
\end{eqnarray*}
Notice that with this choice the coincidence planes are mapped into the nicely symmetric form
\begin{equation}\label{eq:right}
s + t =0,\ 1 + s = 0,\ \mbox{and}\ 1 + t = 0.
\end{equation}
These are the boundaries of a right, isosceles triangle with corners at $(s,t) = (-1,-1)$, $(-1,1)$ and $(1,-1)$. The transformation of $\Delta_\Omega$ induced by the coordinate change $(u,v)$ to $(s,t)$ is a lengthy but straightforward and we do not show it here.

A complete, normalized basis for Lebesgue square-integrable functions on this domain bounded by (\ref{eq:right}) is provided by the functions $h_{n,m}(s,t)$:
\begin{eqnarray}
h_{n,m}(s,t) &=& \frac{1}{4} \left( e^{ \frac{i\pi}{2} \{-n(s+1)+m(t-1)\}}\right. \nonumber\\
&& {} - e^{\frac{i\pi}{2}\{-n(s+1)-m(t-1)\}}  + e^{\frac{i\pi}{2}\{n(s+1)-m(t-1)\}}\nonumber\\
&&{}- e^{\frac{i\pi}{2}\{n(s+1)+m(t-1)\}} - e^{\frac{i\pi}{2}\{-m(s+1)+n(t-1)\}}\nonumber\\
&&{}+ e^{\frac{i\pi}{2}\{-m(s+1)-n(t-1)\}} - e^{\frac{i\pi}{2}\{m(s+1)-n(t-1)\}}\nonumber\\
&&\left. {}+ e^{\frac{i \pi}{2}\{m(s+1)+n(t-1)\}} \right),
\end{eqnarray}
where $n$ and $m$ are positive integers and $n<m$. The matrix elements of the transformed spherical Laplacian can be calculated in this basis and then diagonalized to find the spectrum. In our calculations we set $n,m$ to go up to $N_{max}=80$. With this upper
bound the first 300 energies were quite converged up to the second decimal place. We know this because we did convergence analysis from 60-80 and found out that the eigenenergies up to the 2nd decimal place were not changing. The calculation time for $N_{max}=80$ was approximately 10 days on a reasonably powerful desktop computer. If one is interested in higher excited state energies, then one needs to
increase $N_{max}$ in order to get a better precision at the higher end of
the spectrum. 

Additionally, numerical results were compared to the exact algebraic results for the integrable Coxeter sector for several mass families in $A_3$, $C_3$, and $H_3$ to confirm the uncertainty estimates.  And as described in the main text, we compared the level density of the spectrum to the prediction of Weyl's Law in order to establish that all eigenstates were found by this method.


\begin{thebibliography}{93}%
\makeatletter
\providecommand \@ifxundefined [1]{%
 \@ifx{#1\undefined}
}%
\providecommand \@ifnum [1]{%
 \ifnum #1\expandafter \@firstoftwo
 \else \expandafter \@secondoftwo
 \fi
}%
\providecommand \@ifx [1]{%
 \ifx #1\expandafter \@firstoftwo
 \else \expandafter \@secondoftwo
 \fi
}%
\providecommand \natexlab [1]{#1}%
\providecommand \enquote  [1]{``#1''}%
\providecommand \bibnamefont  [1]{#1}%
\providecommand \bibfnamefont [1]{#1}%
\providecommand \citenamefont [1]{#1}%
\providecommand \href@noop [0]{\@secondoftwo}%
\providecommand \href [0]{\begingroup \@sanitize@url \@href}%
\providecommand \@href[1]{\@@startlink{#1}\@@href}%
\providecommand \@@href[1]{\endgroup#1\@@endlink}%
\providecommand \@sanitize@url [0]{\catcode `\\12\catcode `\$12\catcode
  `\&12\catcode `\#12\catcode `\^12\catcode `\_12\catcode `\%12\relax}%
\providecommand \@@startlink[1]{}%
\providecommand \@@endlink[0]{}%
\providecommand \url  [0]{\begingroup\@sanitize@url \@url }%
\providecommand \@url [1]{\endgroup\@href {#1}{\urlprefix }}%
\providecommand \urlprefix  [0]{URL }%
\providecommand \Eprint [0]{\href }%
\providecommand \doibase [0]{http://dx.doi.org/}%
\providecommand \selectlanguage [0]{\@gobble}%
\providecommand \bibinfo  [0]{\@secondoftwo}%
\providecommand \bibfield  [0]{\@secondoftwo}%
\providecommand \translation [1]{[#1]}%
\providecommand \BibitemOpen [0]{}%
\providecommand \bibitemStop [0]{}%
\providecommand \bibitemNoStop [0]{.\EOS\space}%
\providecommand \EOS [0]{\spacefactor3000\relax}%
\providecommand \BibitemShut  [1]{\csname bibitem#1\endcsname}%
\let\auto@bib@innerbib\@empty
\bibitem [{\citenamefont {Albeverio}\ \emph {et~al.}(1988)\citenamefont
  {Albeverio}, \citenamefont {Gesztesy}, \citenamefont {Hoegh-Krohn},\ and\
  \citenamefont {Holden}}]{albeverio_solvable_2012}%
  \BibitemOpen
  \bibfield  {author} {\bibinfo {author} {\bibfnamefont {Sergio}\ \bibnamefont
  {Albeverio}}, \bibinfo {author} {\bibfnamefont {Friedrich}\ \bibnamefont
  {Gesztesy}}, \bibinfo {author} {\bibfnamefont {Raphael}\ \bibnamefont
  {Hoegh-Krohn}}, \ and\ \bibinfo {author} {\bibfnamefont {Helge}\ \bibnamefont
  {Holden}},\ }\href@noop {} {{\selectlanguage {English}\emph {\bibinfo {title}
  {Solvable {Models} in {Quantum} {Mechanics}}}}}\ (\bibinfo  {publisher}
  {Springer-Verlag},\ \bibinfo {address} {Berlin},\ \bibinfo {year}
  {1988})\BibitemShut {NoStop}%
\bibitem [{\citenamefont {Sutherland}(2004)}]{sutherland_beautiful_2004}%
  \BibitemOpen
  \bibfield  {author} {\bibinfo {author} {\bibfnamefont {Bill}\ \bibnamefont
  {Sutherland}},\ }\href@noop {} {{\selectlanguage {English}\emph {\bibinfo
  {title} {Beautiful {Models}: 70 {Years} of {Exactly} {Solved} {Quantum}
  {Many}-{Body} {Problems}}}}}\ (\bibinfo  {publisher} {World Scientific
  Publishing Company},\ \bibinfo {address} {River Edge, N.J},\ \bibinfo {year}
  {2004})\BibitemShut {NoStop}%
\bibitem [{\citenamefont {Gaudin}(2014)}]{gaudin_bethe_2014}%
  \BibitemOpen
  \bibfield  {author} {\bibinfo {author} {\bibfnamefont {Michel}\ \bibnamefont
  {Gaudin}},\ }\href@noop {} {{\selectlanguage {English}\emph {\bibinfo {title}
  {The {Bethe} {Wavefunction}}}}},\ \bibinfo {edition} {js caux, trans.}\ ed.\
  (\bibinfo  {publisher} {Cambridge University Press},\ \bibinfo {address}
  {Cambridge, United Kingdom ; New York},\ \bibinfo {year} {2014})\BibitemShut
  {NoStop}%
\bibitem [{\citenamefont {Lieb}\ and\ \citenamefont
  {Liniger}(1963)}]{lieb_exact_1963}%
  \BibitemOpen
  \bibfield  {author} {\bibinfo {author} {\bibfnamefont {Elliott~H.}\
  \bibnamefont {Lieb}}\ and\ \bibinfo {author} {\bibfnamefont {Werner}\
  \bibnamefont {Liniger}},\ }\bibfield  {title} {\enquote {\bibinfo {title}
  {Exact {Analysis} of an {Interacting} {Bose} {Gas}. {I}. {The} {General}
  {Solution} and the {Ground} {State}},}\ }\href {\doibase
  10.1103/PhysRev.130.1605} {\bibfield  {journal} {\bibinfo  {journal} {Phys.
  Rev.}\ }\textbf {\bibinfo {volume} {130}},\ \bibinfo {pages} {1605--1616}
  (\bibinfo {year} {1963})}\BibitemShut {NoStop}%
\bibitem [{\citenamefont {Girardeau}(1960)}]{girardeau_relationship_1960}%
  \BibitemOpen
  \bibfield  {author} {\bibinfo {author} {\bibfnamefont {M.}~\bibnamefont
  {Girardeau}},\ }\bibfield  {title} {\enquote {\bibinfo {title} {Relationship
  between {Systems} of {Impenetrable} {Bosons} and {Fermions} in {One}
  {Dimension}},}\ }\href {\doibase doi:10.1063/1.1703687} {\bibfield  {journal}
  {\bibinfo  {journal} {Journal of Mathematical Physics}\ }\textbf {\bibinfo
  {volume} {1}},\ \bibinfo {pages} {516--523} (\bibinfo {year}
  {1960})}\BibitemShut {NoStop}%
\bibitem [{\citenamefont {Calogero}(1971)}]{calogero_solution_1971}%
  \BibitemOpen
  \bibfield  {author} {\bibinfo {author} {\bibfnamefont {F.}~\bibnamefont
  {Calogero}},\ }\bibfield  {title} {\enquote {\bibinfo {title} {Solution of
  the {One}‐{Dimensional} {N}‐{Body} {Problems} with {Quadratic} and/or
  {Inversely} {Quadratic} {Pair} {Potentials}},}\ }\href {\doibase
  10.1063/1.1665604} {\bibfield  {journal} {\bibinfo  {journal} {Journal of
  Mathematical Physics}\ }\textbf {\bibinfo {volume} {12}},\ \bibinfo {pages}
  {419--436} (\bibinfo {year} {1971})}\BibitemShut {NoStop}%
\bibitem [{\citenamefont {Calogero}(2008)}]{calogero_calogero-moser_2008}%
  \BibitemOpen
  \bibfield  {author} {\bibinfo {author} {\bibfnamefont {Francesco}\
  \bibnamefont {Calogero}},\ }\bibfield  {title} {{\selectlanguage
  {english}\enquote {\bibinfo {title} {Calogero-{Moser} system},}\ }}\href
  {\doibase 10.4249/scholarpedia.7216} {\bibfield  {journal} {\bibinfo
  {journal} {Scholarpedia}\ }\textbf {\bibinfo {volume} {3}},\ \bibinfo {pages}
  {7216} (\bibinfo {year} {2008})}\BibitemShut {NoStop}%
\bibitem [{\citenamefont {Olshanetsky}\ and\ \citenamefont
  {Perelomov}(1983)}]{olshanetsky_quantum_1983}%
  \BibitemOpen
  \bibfield  {author} {\bibinfo {author} {\bibfnamefont {M.~A.}\ \bibnamefont
  {Olshanetsky}}\ and\ \bibinfo {author} {\bibfnamefont {A.~M.}\ \bibnamefont
  {Perelomov}},\ }\bibfield  {title} {\enquote {\bibinfo {title} {Quantum
  integrable systems related to lie algebras},}\ }\href {\doibase
  10.1016/0370-1573(83)90018-2} {\bibfield  {journal} {\bibinfo  {journal}
  {Physics Reports}\ }\textbf {\bibinfo {volume} {94}},\ \bibinfo {pages}
  {313--404} (\bibinfo {year} {1983})}\BibitemShut {NoStop}%
\bibitem [{\citenamefont {Cazalilla}\ \emph {et~al.}(2011)\citenamefont
  {Cazalilla}, \citenamefont {Citro}, \citenamefont {Giamarchi}, \citenamefont
  {Orignac},\ and\ \citenamefont {Rigol}}]{cazalilla_one_2011}%
  \BibitemOpen
  \bibfield  {author} {\bibinfo {author} {\bibfnamefont {M.~A.}\ \bibnamefont
  {Cazalilla}}, \bibinfo {author} {\bibfnamefont {R.}~\bibnamefont {Citro}},
  \bibinfo {author} {\bibfnamefont {T.}~\bibnamefont {Giamarchi}}, \bibinfo
  {author} {\bibfnamefont {E.}~\bibnamefont {Orignac}}, \ and\ \bibinfo
  {author} {\bibfnamefont {M.}~\bibnamefont {Rigol}},\ }\bibfield  {title}
  {\enquote {\bibinfo {title} {One dimensional bosons: From condensed matter
  systems to ultracold gases},}\ }\href {\doibase 10.1103/RevModPhys.83.1405}
  {\bibfield  {journal} {\bibinfo  {journal} {Rev. Mod. Phys.}\ }\textbf
  {\bibinfo {volume} {83}},\ \bibinfo {pages} {1405--1466} (\bibinfo {year}
  {2011})}\BibitemShut {NoStop}%
\bibitem [{\citenamefont {Guan}\ \emph {et~al.}(2013)\citenamefont {Guan},
  \citenamefont {Batchelor},\ and\ \citenamefont {Lee}}]{guan_fermi_2013}%
  \BibitemOpen
  \bibfield  {author} {\bibinfo {author} {\bibfnamefont {Xi-Wen}\ \bibnamefont
  {Guan}}, \bibinfo {author} {\bibfnamefont {Murray~T.}\ \bibnamefont
  {Batchelor}}, \ and\ \bibinfo {author} {\bibfnamefont {Chaohong}\
  \bibnamefont {Lee}},\ }\bibfield  {title} {\enquote {\bibinfo {title} {Fermi
  gases in one dimension: {From} {Bethe} ansatz to experiments},}\ }\href
  {\doibase 10.1103/RevModPhys.85.1633} {\bibfield  {journal} {\bibinfo
  {journal} {Rev. Mod. Phys.}\ }\textbf {\bibinfo {volume} {85}},\ \bibinfo
  {pages} {1633--1691} (\bibinfo {year} {2013})}\BibitemShut {NoStop}%
\bibitem [{\citenamefont {Olshanii}(1998)}]{olshanii_atomic_1998}%
  \BibitemOpen
  \bibfield  {author} {\bibinfo {author} {\bibfnamefont {M.}~\bibnamefont
  {Olshanii}},\ }\bibfield  {title} {\enquote {\bibinfo {title} {Atomic
  {Scattering} in the {Presence} of an {External} {Confinement} and a {Gas} of
  {Impenetrable} {Bosons}},}\ }\href {\doibase 10.1103/PhysRevLett.81.938}
  {\bibfield  {journal} {\bibinfo  {journal} {Phys. Rev. Lett.}\ }\textbf
  {\bibinfo {volume} {81}},\ \bibinfo {pages} {938--941} (\bibinfo {year}
  {1998})}\BibitemShut {NoStop}%
\bibitem [{\citenamefont {Kinoshita}\ \emph {et~al.}(2006)\citenamefont
  {Kinoshita}, \citenamefont {Wenger},\ and\ \citenamefont
  {Weiss}}]{kinoshita_quantum_2006}%
  \BibitemOpen
  \bibfield  {author} {\bibinfo {author} {\bibfnamefont {Toshiya}\ \bibnamefont
  {Kinoshita}}, \bibinfo {author} {\bibfnamefont {Trevor}\ \bibnamefont
  {Wenger}}, \ and\ \bibinfo {author} {\bibfnamefont {David~S.}\ \bibnamefont
  {Weiss}},\ }\bibfield  {title} {\enquote {\bibinfo {title} {A quantum
  {Newton}'s cradle},}\ }\href {\doibase 10.1038/nature04693} {\bibfield
  {journal} {\bibinfo  {journal} {Nature}\ }\textbf {\bibinfo {volume} {440}},\
  \bibinfo {pages} {900--903} (\bibinfo {year} {2006})}\BibitemShut {NoStop}%
\bibitem [{\citenamefont {Serwane}\ \emph {et~al.}(2011)\citenamefont
  {Serwane}, \citenamefont {Z\"urn}, \citenamefont {Lompe}, \citenamefont
  {Ottenstein}, \citenamefont {Wenz},\ and\ \citenamefont
  {Jochim}}]{serwane_deterministic_2011}%
  \BibitemOpen
  \bibfield  {author} {\bibinfo {author} {\bibfnamefont {F.}~\bibnamefont
  {Serwane}}, \bibinfo {author} {\bibfnamefont {G.}~\bibnamefont {Z\"urn}},
  \bibinfo {author} {\bibfnamefont {T.}~\bibnamefont {Lompe}}, \bibinfo
  {author} {\bibfnamefont {T.~B.}\ \bibnamefont {Ottenstein}}, \bibinfo
  {author} {\bibfnamefont {A.~N.}\ \bibnamefont {Wenz}}, \ and\ \bibinfo
  {author} {\bibfnamefont {S.}~\bibnamefont {Jochim}},\ }\bibfield  {title}
  {\enquote {\bibinfo {title} {Deterministic {Preparation} of a {Tunable}
  {Few}-{Fermion} {System}},}\ }\href {\doibase 10.1126/science.1201351}
  {\bibfield  {journal} {\bibinfo  {journal} {Science}\ }\textbf {\bibinfo
  {volume} {332}},\ \bibinfo {pages} {336--338} (\bibinfo {year}
  {2011})}\BibitemShut {NoStop}%
\bibitem [{\citenamefont {Wenz}\ \emph {et~al.}(2013)\citenamefont {Wenz},
  \citenamefont {Z\"urn}, \citenamefont {Murmann}, \citenamefont {Brouzos},
  \citenamefont {Lompe},\ and\ \citenamefont {Jochim}}]{wenz_few_2013}%
  \BibitemOpen
  \bibfield  {author} {\bibinfo {author} {\bibfnamefont {A.~N.}\ \bibnamefont
  {Wenz}}, \bibinfo {author} {\bibfnamefont {G.}~\bibnamefont {Z\"urn}},
  \bibinfo {author} {\bibfnamefont {S.}~\bibnamefont {Murmann}}, \bibinfo
  {author} {\bibfnamefont {I.}~\bibnamefont {Brouzos}}, \bibinfo {author}
  {\bibfnamefont {T.}~\bibnamefont {Lompe}}, \ and\ \bibinfo {author}
  {\bibfnamefont {S.}~\bibnamefont {Jochim}},\ }\bibfield  {title} {\enquote
  {\bibinfo {title} {From {Few} to {Many}: {Observing} the {Formation} of a
  {Fermi} {Sea} {One} {Atom} at a {Time}},}\ }\href {\doibase
  10.1126/science.1240516} {\bibfield  {journal} {\bibinfo  {journal}
  {Science}\ }\textbf {\bibinfo {volume} {342}},\ \bibinfo {pages} {457--460}
  (\bibinfo {year} {2013})}\BibitemShut {NoStop}%
\bibitem [{\citenamefont {DiVincenzo}\ \emph {et~al.}(2000)\citenamefont
  {DiVincenzo}, \citenamefont {Bacon}, \citenamefont {Kempe}, \citenamefont
  {Burkard},\ and\ \citenamefont {Whaley}}]{divincenzo_universal_2000}%
  \BibitemOpen
  \bibfield  {author} {\bibinfo {author} {\bibfnamefont {D.~P.}\ \bibnamefont
  {DiVincenzo}}, \bibinfo {author} {\bibfnamefont {D.}~\bibnamefont {Bacon}},
  \bibinfo {author} {\bibfnamefont {J.}~\bibnamefont {Kempe}}, \bibinfo
  {author} {\bibfnamefont {G.}~\bibnamefont {Burkard}}, \ and\ \bibinfo
  {author} {\bibfnamefont {K.~B.}\ \bibnamefont {Whaley}},\ }\bibfield  {title}
  {\enquote {\bibinfo {title} {Universal quantum computation with the exchange
  interaction},}\ }\href {\doibase 10.1038/35042541} {\bibfield  {journal}
  {\bibinfo  {journal} {Nature}\ }\textbf {\bibinfo {volume} {408}},\ \bibinfo
  {pages} {339--342} (\bibinfo {year} {2000})}\BibitemShut {NoStop}%
\bibitem [{\citenamefont {Bose}(2003)}]{PhysRevLett.91.207901}%
  \BibitemOpen
  \bibfield  {author} {\bibinfo {author} {\bibfnamefont {Sougato}\ \bibnamefont
  {Bose}},\ }\bibfield  {title} {\enquote {\bibinfo {title} {Quantum
  communication through an unmodulated spin chain},}\ }\href {\doibase
  10.1103/PhysRevLett.91.207901} {\bibfield  {journal} {\bibinfo  {journal}
  {Phys. Rev. Lett.}\ }\textbf {\bibinfo {volume} {91}},\ \bibinfo {pages}
  {207901} (\bibinfo {year} {2003})}\BibitemShut {NoStop}%
\bibitem [{\citenamefont {Volosniev}\ \emph {et~al.}(2015)\citenamefont
  {Volosniev}, \citenamefont {Petrosyan}, \citenamefont {Valiente},
  \citenamefont {Fedorov}, \citenamefont {Jensen},\ and\ \citenamefont
  {Zinner}}]{PhysRevA.91.023620}%
  \BibitemOpen
  \bibfield  {author} {\bibinfo {author} {\bibfnamefont {A.~G.}\ \bibnamefont
  {Volosniev}}, \bibinfo {author} {\bibfnamefont {D.}~\bibnamefont
  {Petrosyan}}, \bibinfo {author} {\bibfnamefont {M.}~\bibnamefont {Valiente}},
  \bibinfo {author} {\bibfnamefont {D.~V.}\ \bibnamefont {Fedorov}}, \bibinfo
  {author} {\bibfnamefont {A.~S.}\ \bibnamefont {Jensen}}, \ and\ \bibinfo
  {author} {\bibfnamefont {N.~T.}\ \bibnamefont {Zinner}},\ }\bibfield  {title}
  {\enquote {\bibinfo {title} {Engineering the dynamics of effective spin-chain
  models for strongly interacting atomic gases},}\ }\href {\doibase
  10.1103/PhysRevA.91.023620} {\bibfield  {journal} {\bibinfo  {journal} {Phys.
  Rev. A}\ }\textbf {\bibinfo {volume} {91}},\ \bibinfo {pages} {023620}
  (\bibinfo {year} {2015})}\BibitemShut {NoStop}%
\bibitem [{Note1()}]{Note1}%
  \BibitemOpen
  \bibinfo {note} {An exception is the free-space CSM model with mass-scaled
  interaction strengths in Ref.~\cite {sen_multispecies_1996}.}\BibitemShut
  {Stop}%
\bibitem [{\citenamefont {Bohigas}({1991})}]{bohigas1991}%
  \BibitemOpen
  \bibfield  {author} {\bibinfo {author} {\bibfnamefont {O.}~\bibnamefont
  {Bohigas}},\ }\bibfield  {title} {\enquote {\bibinfo {title} {Random matrix
  theories and chaotic dynamics},}\ }in\ \href@noop {} {\emph {\bibinfo
  {booktitle} {Chaos and Quantum Physics, {\rm Les Houches, session LII}}}},\
  \bibinfo {editor} {edited by\ \bibinfo {editor} {\bibfnamefont {M.-J.}\
  \bibnamefont {Gianoni}}, \bibinfo {editor} {\bibfnamefont {A.}~\bibnamefont
  {Voros}}, \ and\ \bibinfo {editor} {\bibfnamefont {J.}~\bibnamefont
  {Zinn-Justin}}}\ (\bibinfo  {publisher} {North-Holland},\ \bibinfo {address}
  {Amsterdam},\ \bibinfo {year} {{1991}})\BibitemShut {NoStop}%
\bibitem [{\citenamefont {Whan}(1997)}]{whan_hierarchical_1997}%
  \BibitemOpen
  \bibfield  {author} {\bibinfo {author} {\bibfnamefont {C.~B.}\ \bibnamefont
  {Whan}},\ }\bibfield  {title} {\enquote {\bibinfo {title} {Hierarchical
  level-clustering in two-dimensional harmonic oscillators},}\ }\href {\doibase
  10.1103/PhysRevE.55.R3813} {\bibfield  {journal} {\bibinfo  {journal} {Phys.
  Rev. E}\ }\textbf {\bibinfo {volume} {55}},\ \bibinfo {pages} {R3813--R3816}
  (\bibinfo {year} {1997})}\BibitemShut {NoStop}%
\bibitem [{\citenamefont {Brouzos}\ and\ \citenamefont
  {Foerster}(2014)}]{PhysRevA.89.053623}%
  \BibitemOpen
  \bibfield  {author} {\bibinfo {author} {\bibfnamefont {Ioannis}\ \bibnamefont
  {Brouzos}}\ and\ \bibinfo {author} {\bibfnamefont {Angela}\ \bibnamefont
  {Foerster}},\ }\bibfield  {title} {\enquote {\bibinfo {title} {Trace of
  broken integrability in stationary correlation properties},}\ }\href
  {\doibase 10.1103/PhysRevA.89.053623} {\bibfield  {journal} {\bibinfo
  {journal} {Phys. Rev. A}\ }\textbf {\bibinfo {volume} {89}},\ \bibinfo
  {pages} {053623} (\bibinfo {year} {2014})}\BibitemShut {NoStop}%
\bibitem [{\citenamefont {Rigol}\ \emph {et~al.}(2008)\citenamefont {Rigol},
  \citenamefont {Dunjko},\ and\ \citenamefont
  {Olshanii}}]{rigol_thermalization_2008}%
  \BibitemOpen
  \bibfield  {author} {\bibinfo {author} {\bibfnamefont {Marcos}\ \bibnamefont
  {Rigol}}, \bibinfo {author} {\bibfnamefont {Vanja}\ \bibnamefont {Dunjko}}, \
  and\ \bibinfo {author} {\bibfnamefont {Maxim}\ \bibnamefont {Olshanii}},\
  }\bibfield  {title} {{\selectlanguage {english}\enquote {\bibinfo {title}
  {Thermalization and its mechanism for generic isolated quantum systems},}\
  }}\href {\doibase 10.1038/nature06838} {\bibfield  {journal} {\bibinfo
  {journal} {Nature}\ }\textbf {\bibinfo {volume} {452}},\ \bibinfo {pages}
  {854--858} (\bibinfo {year} {2008})}\BibitemShut {NoStop}%
\bibitem [{\citenamefont {De~Nardis}\ \emph {et~al.}(2014)\citenamefont
  {De~Nardis}, \citenamefont {Wouters}, \citenamefont {Brockmann},\ and\
  \citenamefont {Caux}}]{PhysRevA.89.033601}%
  \BibitemOpen
  \bibfield  {author} {\bibinfo {author} {\bibfnamefont {Jacopo}\ \bibnamefont
  {De~Nardis}}, \bibinfo {author} {\bibfnamefont {Bram}\ \bibnamefont
  {Wouters}}, \bibinfo {author} {\bibfnamefont {Michael}\ \bibnamefont
  {Brockmann}}, \ and\ \bibinfo {author} {\bibfnamefont {Jean-S\'ebastien}\
  \bibnamefont {Caux}},\ }\bibfield  {title} {\enquote {\bibinfo {title}
  {Solution for an interaction quench in the lieb-liniger bose gas},}\ }\href
  {\doibase 10.1103/PhysRevA.89.033601} {\bibfield  {journal} {\bibinfo
  {journal} {Phys. Rev. A}\ }\textbf {\bibinfo {volume} {89}},\ \bibinfo
  {pages} {033601} (\bibinfo {year} {2014})}\BibitemShut {NoStop}%
\bibitem [{\citenamefont {Neill}\ \emph {et~al.}(2016)\citenamefont {Neill},
  \citenamefont {Roushan}, \citenamefont {Fang}, \citenamefont {Chen},
  \citenamefont {Kolodrubetz}, \citenamefont {Chen}, \citenamefont {Megrant},
  \citenamefont {Barends}, \citenamefont {Campbell}, \citenamefont {Chiaro},
  \citenamefont {Dunsworth}, \citenamefont {Jeffrey}, \citenamefont {Kelly},
  \citenamefont {Mutus}, \citenamefont {O'Malley}, \citenamefont {Quintana},
  \citenamefont {Sank}, \citenamefont {Vainsencher}, \citenamefont {Wenner},
  \citenamefont {White}, \citenamefont {Polkovnikov},\ and\ \citenamefont
  {Martinis}}]{neill_ergodic_2016}%
  \BibitemOpen
  \bibfield  {author} {\bibinfo {author} {\bibfnamefont {C.}~\bibnamefont
  {Neill}}, \bibinfo {author} {\bibfnamefont {P.}~\bibnamefont {Roushan}},
  \bibinfo {author} {\bibfnamefont {M.}~\bibnamefont {Fang}}, \bibinfo {author}
  {\bibfnamefont {Y.}~\bibnamefont {Chen}}, \bibinfo {author} {\bibfnamefont
  {M.}~\bibnamefont {Kolodrubetz}}, \bibinfo {author} {\bibfnamefont
  {Z.}~\bibnamefont {Chen}}, \bibinfo {author} {\bibfnamefont {A.}~\bibnamefont
  {Megrant}}, \bibinfo {author} {\bibfnamefont {R.}~\bibnamefont {Barends}},
  \bibinfo {author} {\bibfnamefont {B.}~\bibnamefont {Campbell}}, \bibinfo
  {author} {\bibfnamefont {B.}~\bibnamefont {Chiaro}}, \bibinfo {author}
  {\bibfnamefont {A.}~\bibnamefont {Dunsworth}}, \bibinfo {author}
  {\bibfnamefont {E.}~\bibnamefont {Jeffrey}}, \bibinfo {author} {\bibfnamefont
  {J.}~\bibnamefont {Kelly}}, \bibinfo {author} {\bibfnamefont
  {J.}~\bibnamefont {Mutus}}, \bibinfo {author} {\bibfnamefont {P.~J.~J.}\
  \bibnamefont {O'Malley}}, \bibinfo {author} {\bibfnamefont {C.}~\bibnamefont
  {Quintana}}, \bibinfo {author} {\bibfnamefont {D.}~\bibnamefont {Sank}},
  \bibinfo {author} {\bibfnamefont {A.}~\bibnamefont {Vainsencher}}, \bibinfo
  {author} {\bibfnamefont {J.}~\bibnamefont {Wenner}}, \bibinfo {author}
  {\bibfnamefont {T.~C.}\ \bibnamefont {White}}, \bibinfo {author}
  {\bibfnamefont {A.}~\bibnamefont {Polkovnikov}}, \ and\ \bibinfo {author}
  {\bibfnamefont {J.~M.}\ \bibnamefont {Martinis}},\ }\bibfield  {title}
  {\enquote {\bibinfo {title} {Ergodic dynamics and thermalization in an
  isolated quantum system},}\ }\href {\doibase 10.1038/nphys3830} {\bibfield
  {journal} {\bibinfo  {journal} {Nat Phys}\ }\textbf {\bibinfo {volume}
  {12}},\ \bibinfo {pages} {1037--1041} (\bibinfo {year} {2016})}\BibitemShut
  {NoStop}%
\bibitem [{\citenamefont {Mehta}(2014)}]{mehta_born-oppenheimer_2014}%
  \BibitemOpen
  \bibfield  {author} {\bibinfo {author} {\bibfnamefont {N.~P.}\ \bibnamefont
  {Mehta}},\ }\bibfield  {title} {\enquote {\bibinfo {title}
  {Born-{Oppenheimer} study of two-component few-particle systems under
  one-dimensional confinement},}\ }\href {\doibase 10.1103/PhysRevA.89.052706}
  {\bibfield  {journal} {\bibinfo  {journal} {Phys. Rev. A}\ }\textbf {\bibinfo
  {volume} {89}},\ \bibinfo {pages} {052706} (\bibinfo {year} {2014})},\
  \bibinfo {note} {[Erratum], ibid 89, 052706}\BibitemShut {NoStop}%
\bibitem [{\citenamefont {Dehkharghani}\ \emph {et~al.}(2015)\citenamefont
  {Dehkharghani}, \citenamefont {Volosniev},\ and\ \citenamefont
  {Zinner}}]{dehkharghani_quantum_2015}%
  \BibitemOpen
  \bibfield  {author} {\bibinfo {author} {\bibfnamefont {A.~S.}\ \bibnamefont
  {Dehkharghani}}, \bibinfo {author} {\bibfnamefont {A.~G.}\ \bibnamefont
  {Volosniev}}, \ and\ \bibinfo {author} {\bibfnamefont {N.~T.}\ \bibnamefont
  {Zinner}},\ }\bibfield  {title} {\enquote {\bibinfo {title} {Quantum impurity
  in a one-dimensional trapped {Bose} gas},}\ }\href {\doibase
  10.1103/PhysRevA.92.031601} {\bibfield  {journal} {\bibinfo  {journal} {Phys.
  Rev. A}\ }\textbf {\bibinfo {volume} {92}},\ \bibinfo {pages} {031601}
  (\bibinfo {year} {2015})}\BibitemShut {NoStop}%
\bibitem [{\citenamefont {P\c{e}cak}\ \emph {et~al.}(2016)\citenamefont
  {P\c{e}cak}, \citenamefont {Gajda},\ and\ \citenamefont
  {Sowi\'{n}ski}}]{pecak_two-flavour_2016}%
  \BibitemOpen
  \bibfield  {author} {\bibinfo {author} {\bibfnamefont {Daniel}\ \bibnamefont
  {P\c{e}cak}}, \bibinfo {author} {\bibfnamefont {Mariusz}\ \bibnamefont
  {Gajda}}, \ and\ \bibinfo {author} {\bibfnamefont {Tomasz}\ \bibnamefont
  {Sowi\'{n}ski}},\ }\bibfield  {title} {{\selectlanguage {english}\enquote
  {\bibinfo {title} {Two-flavour mixture of a few fermions of different mass in
  a one-dimensional harmonic trap},}\ }}\href {\doibase
  10.1088/1367-2630/18/1/013030} {\bibfield  {journal} {\bibinfo  {journal}
  {New J. Phys.}\ }\textbf {\bibinfo {volume} {18}},\ \bibinfo {pages} {013030}
  (\bibinfo {year} {2016})}\BibitemShut {NoStop}%
\bibitem [{\citenamefont {Mehta}\ and\ \citenamefont
  {Morehead}(2015)}]{mehta_few-boson_2015}%
  \BibitemOpen
  \bibfield  {author} {\bibinfo {author} {\bibfnamefont {N.~P.}\ \bibnamefont
  {Mehta}}\ and\ \bibinfo {author} {\bibfnamefont {Connor~D.}\ \bibnamefont
  {Morehead}},\ }\bibfield  {title} {\enquote {\bibinfo {title} {Few-boson
  processes in the presence of an attractive impurity under one-dimensional
  confinement},}\ }\href {\doibase 10.1103/PhysRevA.92.043616} {\bibfield
  {journal} {\bibinfo  {journal} {Phys. Rev. A}\ }\textbf {\bibinfo {volume}
  {92}},\ \bibinfo {pages} {043616} (\bibinfo {year} {2015})}\BibitemShut
  {NoStop}%
\bibitem [{\citenamefont {P\c{e}cak}\ and\ \citenamefont
  {Sowi\'{n}ski}(2016)}]{PhysRevA.94.042118}%
  \BibitemOpen
  \bibfield  {author} {\bibinfo {author} {\bibfnamefont {Daniel}\ \bibnamefont
  {P\c{e}cak}}\ and\ \bibinfo {author} {\bibfnamefont {Tomasz}\ \bibnamefont
  {Sowi\'{n}ski}},\ }\bibfield  {title} {\enquote {\bibinfo {title} {Few
  strongly interacting ultracold fermions in one-dimensional traps of different
  shapes},}\ }\href {\doibase 10.1103/PhysRevA.94.042118} {\bibfield  {journal}
  {\bibinfo  {journal} {Phys. Rev. A}\ }\textbf {\bibinfo {volume} {94}},\
  \bibinfo {pages} {042118} (\bibinfo {year} {2016})}\BibitemShut {NoStop}%
\bibitem [{\citenamefont {Dehkharghani}\ \emph {et~al.}(2016)\citenamefont
  {Dehkharghani}, \citenamefont {Volosniev},\ and\ \citenamefont
  {Zinner}}]{dehkharghani_impenetrable_2016}%
  \BibitemOpen
  \bibfield  {author} {\bibinfo {author} {\bibfnamefont {A.~S.}\ \bibnamefont
  {Dehkharghani}}, \bibinfo {author} {\bibfnamefont {A.~G.}\ \bibnamefont
  {Volosniev}}, \ and\ \bibinfo {author} {\bibfnamefont {N.~T.}\ \bibnamefont
  {Zinner}},\ }\bibfield  {title} {{\selectlanguage {english}\enquote {\bibinfo
  {title} {Impenetrable mass-imbalanced particles in one-dimensional harmonic
  traps},}\ }}\href {\doibase 10.1088/0953-4075/49/8/085301} {\bibfield
  {journal} {\bibinfo  {journal} {J. Phys. B: At. Mol. Opt. Phys.}\ }\textbf
  {\bibinfo {volume} {49}},\ \bibinfo {pages} {085301} (\bibinfo {year}
  {2016})}\BibitemShut {NoStop}%
\bibitem [{\citenamefont {P\c{e}cak}\ \emph
  {et~al.}(2017{\natexlab{a}})\citenamefont {P\c{e}cak}, \citenamefont
  {Gajda},\ and\ \citenamefont {Sowi\'{n}ski}}]{pecak_experimentally_2017}%
  \BibitemOpen
  \bibfield  {author} {\bibinfo {author} {\bibfnamefont {Daniel}\ \bibnamefont
  {P\c{e}cak}}, \bibinfo {author} {\bibfnamefont {Mariusz}\ \bibnamefont
  {Gajda}}, \ and\ \bibinfo {author} {\bibfnamefont {Tomasz}\ \bibnamefont
  {Sowi\'{n}ski}},\ }\bibfield  {title} {\enquote {\bibinfo {title}
  {Experimentally accessible invariants encoded in interparticle correlations
  of harmonically trapped ultra-cold few-fermion mixtures},}\ }\href
  {http://arxiv.org/abs/1703.08116} {\  (\bibinfo {year}
  {2017}{\natexlab{a}})},\ \bibinfo {note} {arXiv: 1703.08116}\BibitemShut
  {NoStop}%
\bibitem [{\citenamefont {P\c{e}cak}\ \emph
  {et~al.}(2017{\natexlab{b}})\citenamefont {P\c{e}cak}, \citenamefont
  {Dehkharghani}, \citenamefont {Zinner},\ and\ \citenamefont
  {Sowi\'{n}ski}}]{PhysRevA.95.053632}%
  \BibitemOpen
  \bibfield  {author} {\bibinfo {author} {\bibfnamefont {D.}~\bibnamefont
  {P\c{e}cak}}, \bibinfo {author} {\bibfnamefont {A.~S.}\ \bibnamefont
  {Dehkharghani}}, \bibinfo {author} {\bibfnamefont {N.~T.}\ \bibnamefont
  {Zinner}}, \ and\ \bibinfo {author} {\bibfnamefont {T.}~\bibnamefont
  {Sowi\'{n}ski}},\ }\bibfield  {title} {\enquote {\bibinfo {title} {Four
  fermions in a one-dimensional harmonic trap: Accuracy of a variational-ansatz
  approach},}\ }\href {\doibase 10.1103/PhysRevA.95.053632} {\bibfield
  {journal} {\bibinfo  {journal} {Phys. Rev. A}\ }\textbf {\bibinfo {volume}
  {95}},\ \bibinfo {pages} {053632} (\bibinfo {year}
  {2017}{\natexlab{b}})}\BibitemShut {NoStop}%
\bibitem [{\citenamefont {Olshanii}\ and\ \citenamefont
  {Jackson}(2015)}]{olshanii_exactly_2015}%
  \BibitemOpen
  \bibfield  {author} {\bibinfo {author} {\bibfnamefont {Maxim}\ \bibnamefont
  {Olshanii}}\ and\ \bibinfo {author} {\bibfnamefont {Steven~G.}\ \bibnamefont
  {Jackson}},\ }\bibfield  {title} {{\selectlanguage {english}\enquote
  {\bibinfo {title} {An exactly solvable quantum four-body problem associated
  with the symmetries of an octacube},}\ }}\href {\doibase
  10.1088/1367-2630/17/10/105005} {\bibfield  {journal} {\bibinfo  {journal}
  {New J. Phys.}\ }\textbf {\bibinfo {volume} {17}},\ \bibinfo {pages} {105005}
  (\bibinfo {year} {2015})}\BibitemShut {NoStop}%
\bibitem [{\citenamefont {Olshanii}\ \emph
  {et~al.}(2016{\natexlab{a}})\citenamefont {Olshanii}, \citenamefont
  {Scoquart}, \citenamefont {Seaward},\ and\ \citenamefont
  {Jackson}}]{olshanii_exactly_2016}%
  \BibitemOpen
  \bibfield  {author} {\bibinfo {author} {\bibfnamefont {Maxim}\ \bibnamefont
  {Olshanii}}, \bibinfo {author} {\bibfnamefont {Thibault}\ \bibnamefont
  {Scoquart}}, \bibinfo {author} {\bibfnamefont {Joseph}\ \bibnamefont
  {Seaward}}, \ and\ \bibinfo {author} {\bibfnamefont {Steven~Glenn}\
  \bibnamefont {Jackson}},\ }\bibfield  {title} {\enquote {\bibinfo {title}
  {Exactly solvable quantum few-body systems associated with the symmetries of
  the three-dimensional and four-dimensional icosahedra},}\ }\href {\doibase
  10.21468/SciPostPhys.1.1.005} {\bibfield  {journal} {\bibinfo  {journal}
  {SciPost Physics}\ }\textbf {\bibinfo {volume} {1}},\ \bibinfo {pages} {005}
  (\bibinfo {year} {2016}{\natexlab{a}})}\BibitemShut {NoStop}%
\bibitem [{\citenamefont {Olshanii}\ \emph
  {et~al.}(2016{\natexlab{b}})\citenamefont {Olshanii}, \citenamefont
  {Scoquart}, \citenamefont {Yampolsky}, \citenamefont {Dunjko},\ and\
  \citenamefont {Jackson}}]{olshanii_creating_2016}%
  \BibitemOpen
  \bibfield  {author} {\bibinfo {author} {\bibfnamefont {Maxim}\ \bibnamefont
  {Olshanii}}, \bibinfo {author} {\bibfnamefont {Thibault}\ \bibnamefont
  {Scoquart}}, \bibinfo {author} {\bibfnamefont {Dmitry}\ \bibnamefont
  {Yampolsky}}, \bibinfo {author} {\bibfnamefont {Vanja}\ \bibnamefont
  {Dunjko}}, \ and\ \bibinfo {author} {\bibfnamefont {Steven~Glenn}\
  \bibnamefont {Jackson}},\ }\bibfield  {title} {\enquote {\bibinfo {title}
  {Creating {Entanglement} {Using} {Integrals} of {Motion}},}\ }\href
  {http://arxiv.org/abs/1610.01060} {\  (\bibinfo {year}
  {2016}{\natexlab{b}})},\ \bibinfo {note} {arXiv: 1610.01060}\BibitemShut
  {NoStop}%
\bibitem [{\citenamefont {Volosniev}\ \emph {et~al.}(2014)\citenamefont
  {Volosniev}, \citenamefont {Fedorov}, \citenamefont {Jensen}, \citenamefont
  {Valiente},\ and\ \citenamefont {Zinner}}]{volosniev_strongly_2014}%
  \BibitemOpen
  \bibfield  {author} {\bibinfo {author} {\bibfnamefont {A.~G.}\ \bibnamefont
  {Volosniev}}, \bibinfo {author} {\bibfnamefont {D.~V.}\ \bibnamefont
  {Fedorov}}, \bibinfo {author} {\bibfnamefont {A.~S.}\ \bibnamefont {Jensen}},
  \bibinfo {author} {\bibfnamefont {M.}~\bibnamefont {Valiente}}, \ and\
  \bibinfo {author} {\bibfnamefont {N.~T.}\ \bibnamefont {Zinner}},\ }\bibfield
   {title} {\enquote {\bibinfo {title} {Strongly interacting confined quantum
  systems in one dimension},}\ }\href {http://dx.doi.org/10.1038/ncomms6300}
  {\bibfield  {journal} {\bibinfo  {journal} {Nature Communications}\ }\textbf
  {\bibinfo {volume} {5}},\ \bibinfo {pages} {5300} (\bibinfo {year}
  {2014})}\BibitemShut {NoStop}%
\bibitem [{\citenamefont {Deuretzbacher}\ \emph {et~al.}(2014)\citenamefont
  {Deuretzbacher}, \citenamefont {Becker}, \citenamefont {Bjerlin},
  \citenamefont {Reimann},\ and\ \citenamefont {Santos}}]{PhysRevA.90.013611}%
  \BibitemOpen
  \bibfield  {author} {\bibinfo {author} {\bibfnamefont {F.}~\bibnamefont
  {Deuretzbacher}}, \bibinfo {author} {\bibfnamefont {D.}~\bibnamefont
  {Becker}}, \bibinfo {author} {\bibfnamefont {J.}~\bibnamefont {Bjerlin}},
  \bibinfo {author} {\bibfnamefont {S.~M.}\ \bibnamefont {Reimann}}, \ and\
  \bibinfo {author} {\bibfnamefont {L.}~\bibnamefont {Santos}},\ }\bibfield
  {title} {\enquote {\bibinfo {title} {Quantum magnetism without lattices in
  strongly interacting one-dimensional spinor gases},}\ }\href {\doibase
  10.1103/PhysRevA.90.013611} {\bibfield  {journal} {\bibinfo  {journal} {Phys.
  Rev. A}\ }\textbf {\bibinfo {volume} {90}},\ \bibinfo {pages} {013611}
  (\bibinfo {year} {2014})}\BibitemShut {NoStop}%
\bibitem [{\citenamefont {Harshman}(2016)}]{Harshman2016}%
  \BibitemOpen
  \bibfield  {author} {\bibinfo {author} {\bibfnamefont {N.~L.}\ \bibnamefont
  {Harshman}},\ }\bibfield  {title} {\enquote {\bibinfo {title}
  {One-dimensional traps, two-body interactions, few-body symmetries. {II}.
  {$N$} particles},}\ }\href {\doibase 10.1007/s00601-015-1025-5} {\bibfield
  {journal} {\bibinfo  {journal} {Few-Body Systems}\ }\textbf {\bibinfo
  {volume} {57}},\ \bibinfo {pages} {45--69} (\bibinfo {year}
  {2016})}\BibitemShut {NoStop}%
\bibitem [{\citenamefont {Roscher}\ \emph {et~al.}(2014)\citenamefont
  {Roscher}, \citenamefont {Braun}, \citenamefont {Chen},\ and\ \citenamefont
  {Drut}}]{0954-3899-41-5-055110}%
  \BibitemOpen
  \bibfield  {author} {\bibinfo {author} {\bibfnamefont {Dietrich}\
  \bibnamefont {Roscher}}, \bibinfo {author} {\bibfnamefont {Jens}\
  \bibnamefont {Braun}}, \bibinfo {author} {\bibfnamefont {Jiunn-Wei}\
  \bibnamefont {Chen}}, \ and\ \bibinfo {author} {\bibfnamefont {Joaquín~E}\
  \bibnamefont {Drut}},\ }\bibfield  {title} {\enquote {\bibinfo {title} {Fermi
  gases with imaginary mass imbalance and the sign problem in monte-carlo
  calculations},}\ }\href {http://stacks.iop.org/0954-3899/41/i=5/a=055110}
  {\bibfield  {journal} {\bibinfo  {journal} {Journal of Physics G: Nuclear and
  Particle Physics}\ }\textbf {\bibinfo {volume} {41}},\ \bibinfo {pages}
  {055110} (\bibinfo {year} {2014})}\BibitemShut {NoStop}%
\bibitem [{\citenamefont {Coxeter}(1973)}]{coxeter_regular_1973}%
  \BibitemOpen
  \bibfield  {author} {\bibinfo {author} {\bibfnamefont {H.~S.~M.}\
  \bibnamefont {Coxeter}},\ }\href@noop {} {{\selectlanguage {English}\emph
  {\bibinfo {title} {Regular {Polytopes}}}}},\ \bibinfo {edition} {3rd}\ ed.\
  (\bibinfo  {publisher} {Dover Publications},\ \bibinfo {address} {New York},\
  \bibinfo {year} {1973})\BibitemShut {NoStop}%
\bibitem [{\citenamefont {Boreskov}\ \emph {et~al.}(2005)\citenamefont
  {Boreskov}, \citenamefont {Turbiner},\ and\ \citenamefont
  {Vieyra}}]{boreskov_solvability_2005}%
  \BibitemOpen
  \bibfield  {author} {\bibinfo {author} {\bibfnamefont {Konstantin~G.}\
  \bibnamefont {Boreskov}}, \bibinfo {author} {\bibfnamefont {Alexander~V.}\
  \bibnamefont {Turbiner}}, \ and\ \bibinfo {author} {\bibfnamefont {Juan
  Carlos~Lopez}\ \bibnamefont {Vieyra}},\ }\bibfield  {title} {{\selectlanguage
  {english}\enquote {\bibinfo {title} {Solvability of the {Hamiltonians}
  {Related} to {Exceptional} {Root} {Spaces}: {Rational} {Case}},}\ }}\href
  {\doibase 10.1007/s00220-005-1401-y} {\bibfield  {journal} {\bibinfo
  {journal} {Commun. Math. Phys.}\ }\textbf {\bibinfo {volume} {260}},\
  \bibinfo {pages} {17--44} (\bibinfo {year} {2005})}\BibitemShut {NoStop}%
\bibitem [{\citenamefont {Garc\'{i}a}\ and\ \citenamefont
  {Turbiner}(2010)}]{garcia_quantum_2010}%
  \BibitemOpen
  \bibfield  {author} {\bibinfo {author} {\bibfnamefont {Marcos A.~G.}\
  \bibnamefont {Garc\'{i}a}}\ and\ \bibinfo {author} {\bibfnamefont
  {Alexander~V.}\ \bibnamefont {Turbiner}},\ }\bibfield  {title} {\enquote
  {\bibinfo {title} {The quantum $\mathrm{H}_3$ integrable system},}\ }\href
  {\doibase 10.1142/S0217751X10050597} {\bibfield  {journal} {\bibinfo
  {journal} {Int. J. Mod. Phys. A}\ }\textbf {\bibinfo {volume} {25}},\
  \bibinfo {pages} {5567--5594} (\bibinfo {year} {2010})}\BibitemShut {NoStop}%
\bibitem [{\citenamefont {Garc\'{i}a}\ and\ \citenamefont
  {Turbiner}(2011)}]{garcia_quantum_2011}%
  \BibitemOpen
  \bibfield  {author} {\bibinfo {author} {\bibfnamefont {Marcos A.~G.}\
  \bibnamefont {Garc\'{i}a}}\ and\ \bibinfo {author} {\bibfnamefont
  {Alexander~V.}\ \bibnamefont {Turbiner}},\ }\bibfield  {title} {\enquote
  {\bibinfo {title} {The quantum $\mathrm{H}_4$ integrable system},}\ }\href
  {\doibase 10.1142/S0217732311034839} {\bibfield  {journal} {\bibinfo
  {journal} {Mod. Phys. Lett. A}\ }\textbf {\bibinfo {volume} {26}},\ \bibinfo
  {pages} {433--447} (\bibinfo {year} {2011})}\BibitemShut {NoStop}%
\bibitem [{\citenamefont {Tempesta}\ \emph {et~al.}(2001)\citenamefont
  {Tempesta}, \citenamefont {Turbiner},\ and\ \citenamefont
  {Winternitz}}]{tempesta_exact_2001}%
  \BibitemOpen
  \bibfield  {author} {\bibinfo {author} {\bibfnamefont {Piergiulio}\
  \bibnamefont {Tempesta}}, \bibinfo {author} {\bibfnamefont {Alexander~V.}\
  \bibnamefont {Turbiner}}, \ and\ \bibinfo {author} {\bibfnamefont {Pavel}\
  \bibnamefont {Winternitz}},\ }\bibfield  {title} {\enquote {\bibinfo {title}
  {Exact solvability of superintegrable systems},}\ }\href {\doibase
  10.1063/1.1386927} {\bibfield  {journal} {\bibinfo  {journal} {Journal of
  Mathematical Physics}\ }\textbf {\bibinfo {volume} {42}},\ \bibinfo {pages}
  {4248--4257} (\bibinfo {year} {2001})}\BibitemShut {NoStop}%
\bibitem [{\citenamefont {Post}\ \emph {et~al.}(2012)\citenamefont {Post},
  \citenamefont {Tsujimoto},\ and\ \citenamefont {Vinet}}]{post_families_2012}%
  \BibitemOpen
  \bibfield  {author} {\bibinfo {author} {\bibfnamefont {Sarah}\ \bibnamefont
  {Post}}, \bibinfo {author} {\bibfnamefont {Satoshi}\ \bibnamefont
  {Tsujimoto}}, \ and\ \bibinfo {author} {\bibfnamefont {Luc}\ \bibnamefont
  {Vinet}},\ }\bibfield  {title} {\enquote {\bibinfo {title} {Families of
  superintegrable {Hamiltonians} constructed from exceptional polynomials},}\
  }\href {\doibase 10.1088/1751-8113/45/40/405202} {\bibfield  {journal}
  {\bibinfo  {journal} {J. Phys. A: Math. Theo.}\ }\textbf {\bibinfo {volume}
  {45}},\ \bibinfo {pages} {405202} (\bibinfo {year} {2012})}\BibitemShut
  {NoStop}%
\bibitem [{\citenamefont {McGuire}(1964)}]{mcguire_study_1964}%
  \BibitemOpen
  \bibfield  {author} {\bibinfo {author} {\bibfnamefont {J.~B.}\ \bibnamefont
  {McGuire}},\ }\bibfield  {title} {\enquote {\bibinfo {title} {Study of
  {Exactly} {Soluble} {One}‐{Dimensional} {N}‐{Body} {Problems}},}\ }\href
  {\doibase doi:10.1063/1.1704156} {\bibfield  {journal} {\bibinfo  {journal}
  {J. Math. Phys.}\ }\textbf {\bibinfo {volume} {5}},\ \bibinfo {pages}
  {622--636} (\bibinfo {year} {1964})}\BibitemShut {NoStop}%
\bibitem [{\citenamefont {Sutherland}(1980)}]{sutherland_nondiffractive_1980}%
  \BibitemOpen
  \bibfield  {author} {\bibinfo {author} {\bibfnamefont {Bill}\ \bibnamefont
  {Sutherland}},\ }\bibfield  {title} {\enquote {\bibinfo {title}
  {Nondiffractive scattering: {Scattering} from kaleidoscopes},}\ }\href
  {\doibase 10.1063/1.524628} {\bibfield  {journal} {\bibinfo  {journal}
  {Journal of Mathematical Physics}\ }\textbf {\bibinfo {volume} {21}},\
  \bibinfo {pages} {1770--1775} (\bibinfo {year} {1980})}\BibitemShut {NoStop}%
\bibitem [{\citenamefont {Lamacraft}(2013)}]{lamacraft_diffractive_2013}%
  \BibitemOpen
  \bibfield  {author} {\bibinfo {author} {\bibfnamefont {Austen}\ \bibnamefont
  {Lamacraft}},\ }\bibfield  {title} {\enquote {\bibinfo {title} {Diffractive
  scattering of three particles in one dimension: {A} simple result for weak
  violations of the {Yang}-{Baxter} equation},}\ }\href {\doibase
  10.1103/PhysRevA.87.012707} {\bibfield  {journal} {\bibinfo  {journal} {Phys.
  Rev. A}\ }\textbf {\bibinfo {volume} {87}},\ \bibinfo {pages} {012707}
  (\bibinfo {year} {2013})}\BibitemShut {NoStop}%
\bibitem [{\citenamefont {Caux}\ and\ \citenamefont
  {Mossel}(2011)}]{caux_remarks_2011}%
  \BibitemOpen
  \bibfield  {author} {\bibinfo {author} {\bibfnamefont {Jean-S\'{e}bastien}\
  \bibnamefont {Caux}}\ and\ \bibinfo {author} {\bibfnamefont {Jorn}\
  \bibnamefont {Mossel}},\ }\bibfield  {title} {{\selectlanguage
  {english}\enquote {\bibinfo {title} {Remarks on the notion of quantum
  integrability},}\ }}\href {\doibase 10.1088/1742-5468/2011/02/P02023}
  {\bibfield  {journal} {\bibinfo  {journal} {J. Stat. Mech.}\ }\textbf
  {\bibinfo {volume} {2011}},\ \bibinfo {pages} {P02023} (\bibinfo {year}
  {2011})}\BibitemShut {NoStop}%
\bibitem [{\citenamefont {Evans}(1990)}]{evans_superintegrability_1990}%
  \BibitemOpen
  \bibfield  {author} {\bibinfo {author} {\bibfnamefont {N.~W.}\ \bibnamefont
  {Evans}},\ }\bibfield  {title} {\enquote {\bibinfo {title}
  {Superintegrability in classical mechanics},}\ }\href {\doibase
  10.1103/PhysRevA.41.5666} {\bibfield  {journal} {\bibinfo  {journal} {Phys.
  Rev. A}\ }\textbf {\bibinfo {volume} {41}},\ \bibinfo {pages} {5666--5676}
  (\bibinfo {year} {1990})}\BibitemShut {NoStop}%
\bibitem [{\citenamefont
  {Wojciechowski}(1983)}]{wojciechowski_superintegrability_1983}%
  \BibitemOpen
  \bibfield  {author} {\bibinfo {author} {\bibfnamefont {S.}~\bibnamefont
  {Wojciechowski}},\ }\bibfield  {title} {\enquote {\bibinfo {title}
  {Superintegrability of the {Calogero}-{Moser} system},}\ }\href {\doibase
  10.1016/0375-9601(83)90018-X} {\bibfield  {journal} {\bibinfo  {journal}
  {Physics Letters A}\ }\textbf {\bibinfo {volume} {95}},\ \bibinfo {pages}
  {279--281} (\bibinfo {year} {1983})}\BibitemShut {NoStop}%
\bibitem [{\citenamefont {Hakobyan}\ \emph {et~al.}(2014)\citenamefont
  {Hakobyan}, \citenamefont {Lechtenfeld},\ and\ \citenamefont
  {Nersessian}}]{hakobyan_superintegrability_2014}%
  \BibitemOpen
  \bibfield  {author} {\bibinfo {author} {\bibfnamefont {Tigran}\ \bibnamefont
  {Hakobyan}}, \bibinfo {author} {\bibfnamefont {Olaf}\ \bibnamefont
  {Lechtenfeld}}, \ and\ \bibinfo {author} {\bibfnamefont {Armen}\ \bibnamefont
  {Nersessian}},\ }\bibfield  {title} {\enquote {\bibinfo {title}
  {Superintegrability of generalized {Calogero} models with oscillator or
  {Coulomb} potential},}\ }\href {\doibase 10.1103/PhysRevD.90.101701}
  {\bibfield  {journal} {\bibinfo  {journal} {Phys. Rev. D}\ }\textbf {\bibinfo
  {volume} {90}},\ \bibinfo {pages} {101701} (\bibinfo {year}
  {2014})}\BibitemShut {NoStop}%
\bibitem [{\citenamefont {Casati}\ and\ \citenamefont
  {Ford}(1976)}]{casati_computer_1976}%
  \BibitemOpen
  \bibfield  {author} {\bibinfo {author} {\bibfnamefont {Giulio}\ \bibnamefont
  {Casati}}\ and\ \bibinfo {author} {\bibfnamefont {Joseph}\ \bibnamefont
  {Ford}},\ }\bibfield  {title} {\enquote {\bibinfo {title} {Computer study of
  ergodicity and mixing in a two-particle, hard point gas system},}\ }\href
  {\doibase 10.1016/0021-9991(76)90104-2} {\bibfield  {journal} {\bibinfo
  {journal} {Journal of Computational Physics}\ }\textbf {\bibinfo {volume}
  {20}},\ \bibinfo {pages} {97--109} (\bibinfo {year} {1976})}\BibitemShut
  {NoStop}%
\bibitem [{\citenamefont {Richens}\ and\ \citenamefont
  {Berry}(1981)}]{richens_pseudointegrable_1981}%
  \BibitemOpen
  \bibfield  {author} {\bibinfo {author} {\bibfnamefont {P.~J.}\ \bibnamefont
  {Richens}}\ and\ \bibinfo {author} {\bibfnamefont {M.~V.}\ \bibnamefont
  {Berry}},\ }\bibfield  {title} {\enquote {\bibinfo {title} {Pseudointegrable
  systems in classical and quantum mechanics},}\ }\href {\doibase
  10.1016/0167-2789(81)90024-5} {\bibfield  {journal} {\bibinfo  {journal}
  {Physica D: Nonlinear Phenomena}\ }\textbf {\bibinfo {volume} {2}},\ \bibinfo
  {pages} {495--512} (\bibinfo {year} {1981})}\BibitemShut {NoStop}%
\bibitem [{\citenamefont {Gutkin}(1996)}]{gutkin_billiards_1996}%
  \BibitemOpen
  \bibfield  {author} {\bibinfo {author} {\bibfnamefont {Eugene}\ \bibnamefont
  {Gutkin}},\ }\bibfield  {title} {{\selectlanguage {english}\enquote {\bibinfo
  {title} {Billiards in polygons: {Survey} of recent results},}\ }}\href
  {\doibase 10.1007/BF02183637} {\bibfield  {journal} {\bibinfo  {journal} {J
  Stat Phys}\ }\textbf {\bibinfo {volume} {83}},\ \bibinfo {pages} {7--26}
  (\bibinfo {year} {1996})}\BibitemShut {NoStop}%
\bibitem [{\citenamefont {Glashow}\ and\ \citenamefont
  {Mittag}(1997)}]{glashow_three_1997}%
  \BibitemOpen
  \bibfield  {author} {\bibinfo {author} {\bibfnamefont {Sheldon~Lee}\
  \bibnamefont {Glashow}}\ and\ \bibinfo {author} {\bibfnamefont {Laurence}\
  \bibnamefont {Mittag}},\ }\bibfield  {title} {{\selectlanguage
  {english}\enquote {\bibinfo {title} {Three rods on a ring and the triangular
  billiard},}\ }}\href {\doibase 10.1007/BF02181254} {\bibfield  {journal}
  {\bibinfo  {journal} {J Stat Phys}\ }\textbf {\bibinfo {volume} {87}},\
  \bibinfo {pages} {937--941} (\bibinfo {year} {1997})}\BibitemShut {NoStop}%
\bibitem [{\citenamefont {Artuso}\ \emph {et~al.}(1997)\citenamefont {Artuso},
  \citenamefont {Casati},\ and\ \citenamefont
  {Guarneri}}]{artuso_numerical_1997}%
  \BibitemOpen
  \bibfield  {author} {\bibinfo {author} {\bibfnamefont {Roberto}\ \bibnamefont
  {Artuso}}, \bibinfo {author} {\bibfnamefont {Giulio}\ \bibnamefont {Casati}},
  \ and\ \bibinfo {author} {\bibfnamefont {Italo}\ \bibnamefont {Guarneri}},\
  }\bibfield  {title} {\enquote {\bibinfo {title} {Numerical study on ergodic
  properties of triangular billiards},}\ }\href {\doibase
  10.1103/PhysRevE.55.6384} {\bibfield  {journal} {\bibinfo  {journal} {Phys.
  Rev. E}\ }\textbf {\bibinfo {volume} {55}},\ \bibinfo {pages} {6384--6390}
  (\bibinfo {year} {1997})}\BibitemShut {NoStop}%
\bibitem [{\citenamefont {Casati}\ and\ \citenamefont
  {Prosen}(1999)}]{casati_mixing_1999}%
  \BibitemOpen
  \bibfield  {author} {\bibinfo {author} {\bibfnamefont {Giulio}\ \bibnamefont
  {Casati}}\ and\ \bibinfo {author} {\bibfnamefont {Tomaž}\ \bibnamefont
  {Prosen}},\ }\bibfield  {title} {\enquote {\bibinfo {title} {Mixing
  {Property} of {Triangular} {Billiards}},}\ }\href {\doibase
  10.1103/PhysRevLett.83.4729} {\bibfield  {journal} {\bibinfo  {journal}
  {Phys. Rev. Lett.}\ }\textbf {\bibinfo {volume} {83}},\ \bibinfo {pages}
  {4729--4732} (\bibinfo {year} {1999})}\BibitemShut {NoStop}%
\bibitem [{\citenamefont {McGuire}\ and\ \citenamefont
  {Dirk}(2001)}]{mcguire_extending_2001}%
  \BibitemOpen
  \bibfield  {author} {\bibinfo {author} {\bibfnamefont {J.~B.}\ \bibnamefont
  {McGuire}}\ and\ \bibinfo {author} {\bibfnamefont {Charlotte}\ \bibnamefont
  {Dirk}},\ }\bibfield  {title} {{\selectlanguage {english}\enquote {\bibinfo
  {title} {Extending the {Bethe} {Ansatz}: {The} {Quantum} {Three}-{Particle}
  {Ring}},}\ }}\href {\doibase 10.1023/A:1004815406443} {\bibfield  {journal}
  {\bibinfo  {journal} {Journal of Statistical Physics}\ }\textbf {\bibinfo
  {volume} {102}},\ \bibinfo {pages} {971--980} (\bibinfo {year}
  {2001})}\BibitemShut {NoStop}%
\bibitem [{\citenamefont {Ara\'{u}jo~Lima}\ \emph {et~al.}(2013)\citenamefont
  {Ara\'{u}jo~Lima}, \citenamefont {Rodr\'{i}guez-P\'{e}rez},\ and\ \citenamefont
  {de~Aguiar}}]{araujo_lima_ergodicity_2013}%
  \BibitemOpen
  \bibfield  {author} {\bibinfo {author} {\bibfnamefont {T.}~\bibnamefont
  {Ara\'{u}jo~Lima}}, \bibinfo {author} {\bibfnamefont {S.}~\bibnamefont
  {Rodr\'{i}guez-P\'{e}rez}}, \ and\ \bibinfo {author} {\bibfnamefont {F.~M.}\
  \bibnamefont {de~Aguiar}},\ }\bibfield  {title} {\enquote {\bibinfo {title}
  {Ergodicity and quantum correlations in irrational triangular billiards},}\
  }\href {\doibase 10.1103/PhysRevE.87.062902} {\bibfield  {journal} {\bibinfo
  {journal} {Phys. Rev. E}\ }\textbf {\bibinfo {volume} {87}},\ \bibinfo
  {pages} {062902} (\bibinfo {year} {2013})}\BibitemShut {NoStop}%
\bibitem [{\citenamefont {Wang}\ \emph {et~al.}(2014)\citenamefont {Wang},
  \citenamefont {Casati},\ and\ \citenamefont
  {Prosen}}]{wang_nonergodicity_2014}%
  \BibitemOpen
  \bibfield  {author} {\bibinfo {author} {\bibfnamefont {Jiao}\ \bibnamefont
  {Wang}}, \bibinfo {author} {\bibfnamefont {Giulio}\ \bibnamefont {Casati}}, \
  and\ \bibinfo {author} {\bibfnamefont {Tomaž}\ \bibnamefont {Prosen}},\
  }\bibfield  {title} {\enquote {\bibinfo {title} {Nonergodicity and
  localization of invariant measure for two colliding masses},}\ }\href
  {\doibase 10.1103/PhysRevE.89.042918} {\bibfield  {journal} {\bibinfo
  {journal} {Phys. Rev. E}\ }\textbf {\bibinfo {volume} {89}},\ \bibinfo
  {pages} {042918} (\bibinfo {year} {2014})}\BibitemShut {NoStop}%
\bibitem [{\citenamefont {Humphreys}(1992)}]{humphreys_reflection_1992}%
  \BibitemOpen
  \bibfield  {author} {\bibinfo {author} {\bibfnamefont {James~E.}\
  \bibnamefont {Humphreys}},\ }\href@noop {} {{\selectlanguage {english}\emph
  {\bibinfo {title} {Reflection {Groups} and {Coxeter} {Groups}}}}}\ (\bibinfo
  {publisher} {Cambridge University Press},\ \bibinfo {year}
  {1992})\BibitemShut {NoStop}%
\bibitem [{\citenamefont {Goodman}(2004)}]{goodman_alice_2004}%
  \BibitemOpen
  \bibfield  {author} {\bibinfo {author} {\bibfnamefont {Roe}\ \bibnamefont
  {Goodman}},\ }\bibfield  {title} {\enquote {\bibinfo {title} {Alice through
  {Looking} {Glass} after {Looking} {Glass}: {The} {Mathematics} of {Mirrors}
  and {Kaleidoscopes}},}\ }\href {\doibase 10.2307/4145238} {\bibfield
  {journal} {\bibinfo  {journal} {The American Mathematical Monthly}\ }\textbf
  {\bibinfo {volume} {111}},\ \bibinfo {pages} {281--298} (\bibinfo {year}
  {2004})}\BibitemShut {NoStop}%
\bibitem [{\citenamefont {Yukalov}\ and\ \citenamefont
  {Girardeau}(2005)}]{yukalov_fermi-bose_2005}%
  \BibitemOpen
  \bibfield  {author} {\bibinfo {author} {\bibfnamefont {V.I.}\ \bibnamefont
  {Yukalov}}\ and\ \bibinfo {author} {\bibfnamefont {M.D.}\ \bibnamefont
  {Girardeau}},\ }\bibfield  {title} {{\selectlanguage {english}\enquote
  {\bibinfo {title} {Fermi-{Bose} mapping for one-dimensional {Bose} gases},}\
  }}\href {\doibase 10.1002/lapl.200510011} {\bibfield  {journal} {\bibinfo
  {journal} {Laser Phys. Lett.}\ }\textbf {\bibinfo {volume} {2}},\ \bibinfo
  {pages} {375--382} (\bibinfo {year} {2005})}\BibitemShut {NoStop}%
\bibitem [{\citenamefont {Harshman}(2014)}]{harshman_spectroscopy_2014}%
  \BibitemOpen
  \bibfield  {author} {\bibinfo {author} {\bibfnamefont {N.~L.}\ \bibnamefont
  {Harshman}},\ }\bibfield  {title} {\enquote {\bibinfo {title} {Spectroscopy
  for a few atoms harmonically trapped in one dimension},}\ }\href {\doibase
  10.1103/PhysRevA.89.033633} {\bibfield  {journal} {\bibinfo  {journal} {Phys.
  Rev. A}\ }\textbf {\bibinfo {volume} {89}},\ \bibinfo {pages} {033633}
  (\bibinfo {year} {2014})}\BibitemShut {NoStop}%
\bibitem [{\citenamefont {Brink}\ \emph {et~al.}(1992)\citenamefont {Brink},
  \citenamefont {Hansson},\ and\ \citenamefont
  {Vasiliev}}]{brink_explicit_1992}%
  \BibitemOpen
  \bibfield  {author} {\bibinfo {author} {\bibfnamefont {L.}~\bibnamefont
  {Brink}}, \bibinfo {author} {\bibfnamefont {T.~H.}\ \bibnamefont {Hansson}},
  \ and\ \bibinfo {author} {\bibfnamefont {M.~A.}\ \bibnamefont {Vasiliev}},\
  }\bibfield  {title} {\enquote {\bibinfo {title} {Explicit solution to the
  {N}-body {Calogero} problem},}\ }\href {\doibase
  10.1016/0370-2693(92)90166-2} {\bibfield  {journal} {\bibinfo  {journal}
  {Phys. Lett. B}\ }\textbf {\bibinfo {volume} {286}},\ \bibinfo {pages}
  {109--111} (\bibinfo {year} {1992})}\BibitemShut {NoStop}%
\bibitem [{\citenamefont {Vacek}\ \emph {et~al.}(1994)\citenamefont {Vacek},
  \citenamefont {Okiji},\ and\ \citenamefont
  {Kawakami}}]{vacek_eigenfunctions_1994}%
  \BibitemOpen
  \bibfield  {author} {\bibinfo {author} {\bibfnamefont {K.}~\bibnamefont
  {Vacek}}, \bibinfo {author} {\bibfnamefont {A.}~\bibnamefont {Okiji}}, \ and\
  \bibinfo {author} {\bibfnamefont {N.}~\bibnamefont {Kawakami}},\ }\bibfield
  {title} {{\selectlanguage {english}\enquote {\bibinfo {title} {Eigenfunctions
  for $\mathrm{SU}( \nu )$ particles with $1/r^2$ interaction in harmonic
  confinement},}\ }}\href {\doibase 10.1088/0305-4470/27/7/002} {\bibfield
  {journal} {\bibinfo  {journal} {J. Phys. A: Math. Gen.}\ }\textbf {\bibinfo
  {volume} {27}},\ \bibinfo {pages} {L201} (\bibinfo {year}
  {1994})}\BibitemShut {NoStop}%
\bibitem [{\citenamefont {Avery}(1989)}]{avery_hyperspherical_1989}%
  \BibitemOpen
  \bibfield  {author} {\bibinfo {author} {\bibfnamefont {John~S.}\ \bibnamefont
  {Avery}},\ }\href@noop {} {{\selectlanguage {english}\emph {\bibinfo {title}
  {Hyperspherical {Harmonics}: {Applications} in {Quantum} {Theory}}}}}\
  (\bibinfo  {publisher} {Springer Netherlands},\ \bibinfo {year}
  {1989})\BibitemShut {NoStop}%
\bibitem [{\citenamefont {Y\'{a}\~{n}ez}\ \emph {et~al.}(1994)\citenamefont
  {Y\'{a}\~{n}ez}, \citenamefont {Van~Assche},\ and\ \citenamefont
  {Dehesa}}]{yanez_position_1994}%
  \BibitemOpen
  \bibfield  {author} {\bibinfo {author} {\bibfnamefont {R.~J.}\ \bibnamefont
  {Y\'{a}\~{n}ez}}, \bibinfo {author} {\bibfnamefont {W.}~\bibnamefont
  {Van~Assche}}, \ and\ \bibinfo {author} {\bibfnamefont {J.~S.}\ \bibnamefont
  {Dehesa}},\ }\bibfield  {title} {\enquote {\bibinfo {title} {Position and
  momentum information entropies of the {D}-dimensional harmonic oscillator and
  hydrogen atom},}\ }\href {\doibase 10.1103/PhysRevA.50.3065} {\bibfield
  {journal} {\bibinfo  {journal} {Phys. Rev. A}\ }\textbf {\bibinfo {volume}
  {50}},\ \bibinfo {pages} {3065--3079} (\bibinfo {year} {1994})}\BibitemShut
  {NoStop}%
\bibitem [{\citenamefont {Ivrii}(2016)}]{ivrii_100_2016}%
  \BibitemOpen
  \bibfield  {author} {\bibinfo {author} {\bibfnamefont {Victor}\ \bibnamefont
  {Ivrii}},\ }\bibfield  {title} {{\selectlanguage {english}\enquote {\bibinfo
  {title} {100 years of {Weyl}’s law},}\ }}\href {\doibase
  10.1007/s13373-016-0089-y} {\bibfield  {journal} {\bibinfo  {journal} {Bull.
  Math. Sci.}\ }\textbf {\bibinfo {volume} {6}},\ \bibinfo {pages} {379--452}
  (\bibinfo {year} {2016})}\BibitemShut {NoStop}%
\bibitem [{\citenamefont {Brooks}\ and\ \citenamefont
  {Strantzen}(2005)}]{brooks_spherical_2005}%
  \BibitemOpen
  \bibfield  {author} {\bibinfo {author} {\bibfnamefont {Jeff}\ \bibnamefont
  {Brooks}}\ and\ \bibinfo {author} {\bibfnamefont {John}\ \bibnamefont
  {Strantzen}},\ }\bibfield  {title} {\enquote {\bibinfo {title} {Spherical
  {Triangles} of {Area} $\pi$ and {Isosceles} {Tetrahedra}},}\ }\href {\doibase
  10.2307/30044179} {\bibfield  {journal} {\bibinfo  {journal} {Mathematics
  Magazine}\ }\textbf {\bibinfo {volume} {78}},\ \bibinfo {pages} {311--314}
  (\bibinfo {year} {2005})}\BibitemShut {NoStop}%
\bibitem [{\citenamefont {Weisstein}(2017)}]{weisstein_spherical_nodate}%
  \BibitemOpen
  \bibfield  {author} {\bibinfo {author} {\bibfnamefont {Eric~W.}\ \bibnamefont
  {Weisstein}},\ }\href
  {http://mathworld.wolfram.com/SphericalTrigonometry.html} {\enquote {\bibinfo
  {title} {Spherical {Trigonometry} -- from {Wolfram} {MathWorld}},}\ }
  (\bibinfo {year} {2017})\BibitemShut {NoStop}%
\bibitem [{\citenamefont {Hobson}(1975)}]{hobson_ergodic_1975}%
  \BibitemOpen
  \bibfield  {author} {\bibinfo {author} {\bibfnamefont {Art}\ \bibnamefont
  {Hobson}},\ }\bibfield  {title} {\enquote {\bibinfo {title} {Ergodic
  properties of a particle moving elastically inside a polygon},}\ }\href
  {\doibase 10.1063/1.522470} {\bibfield  {journal} {\bibinfo  {journal}
  {Journal of Mathematical Physics}\ }\textbf {\bibinfo {volume} {16}},\
  \bibinfo {pages} {2210--2214} (\bibinfo {year} {1975})}\BibitemShut {NoStop}%
\bibitem [{\citenamefont {Mehta}(1988)}]{mehta_basic_1988}%
  \BibitemOpen
  \bibfield  {author} {\bibinfo {author} {\bibfnamefont {M.~L.}\ \bibnamefont
  {Mehta}},\ }\bibfield  {title} {\enquote {\bibinfo {title} {Basic sets of
  invariant polynomials for finite reflection groups},}\ }\href {\doibase
  10.1080/00927878808823619} {\bibfield  {journal} {\bibinfo  {journal}
  {Communications in Algebra}\ }\textbf {\bibinfo {volume} {16}},\ \bibinfo
  {pages} {1083--1098} (\bibinfo {year} {1988})}\BibitemShut {NoStop}%
\bibitem [{\citenamefont {Saghatelian}(2012)}]{saghatelian_constants_2012}%
  \BibitemOpen
  \bibfield  {author} {\bibinfo {author} {\bibfnamefont {A.}~\bibnamefont
  {Saghatelian}},\ }\bibfield  {title} {{\selectlanguage {english}\enquote
  {\bibinfo {title} {Constants of motion of the four-particle {Calogero}
  model},}\ }}\href {\doibase 10.1134/S1063778812100171} {\bibfield  {journal}
  {\bibinfo  {journal} {Phys. Atom. Nuclei}\ }\textbf {\bibinfo {volume}
  {75}},\ \bibinfo {pages} {1288--1293} (\bibinfo {year} {2012})}\BibitemShut
  {NoStop}%
\bibitem [{\citenamefont {Kittel}(2004)}]{kittel_introduction_2004}%
  \BibitemOpen
  \bibfield  {author} {\bibinfo {author} {\bibfnamefont {Charles}\ \bibnamefont
  {Kittel}},\ }\href@noop {} {{\selectlanguage {English}\emph {\bibinfo {title}
  {Introduction to {Solid} {State} {Physics}}}}},\ \bibinfo {edition} {8th}\
  ed.\ (\bibinfo  {publisher} {Wiley},\ \bibinfo {address} {Hoboken, NJ},\
  \bibinfo {year} {2004})\BibitemShut {NoStop}%
\bibitem [{\citenamefont {Gadway}\ \emph {et~al.}(2011)\citenamefont {Gadway},
  \citenamefont {Pertot}, \citenamefont {Reeves}, \citenamefont {Vogt},\ and\
  \citenamefont {Schneble}}]{gadway_glassy_2011}%
  \BibitemOpen
  \bibfield  {author} {\bibinfo {author} {\bibfnamefont {Bryce}\ \bibnamefont
  {Gadway}}, \bibinfo {author} {\bibfnamefont {Daniel}\ \bibnamefont {Pertot}},
  \bibinfo {author} {\bibfnamefont {Jeremy}\ \bibnamefont {Reeves}}, \bibinfo
  {author} {\bibfnamefont {Matthias}\ \bibnamefont {Vogt}}, \ and\ \bibinfo
  {author} {\bibfnamefont {Dominik}\ \bibnamefont {Schneble}},\ }\bibfield
  {title} {\enquote {\bibinfo {title} {Glassy {Behavior} in a {Binary} {Atomic}
  {Mixture}},}\ }\href {\doibase 10.1103/PhysRevLett.107.145306} {\bibfield
  {journal} {\bibinfo  {journal} {Phys. Rev. Lett.}\ }\textbf {\bibinfo
  {volume} {107}},\ \bibinfo {pages} {145306} (\bibinfo {year}
  {2011})}\BibitemShut {NoStop}%
\bibitem [{\citenamefont {Khaykovich}\ \emph {et~al.}(2002)\citenamefont
  {Khaykovich}, \citenamefont {Schreck}, \citenamefont {Ferrari}, \citenamefont
  {Bourdel}, \citenamefont {Cubizolles}, \citenamefont {Carr}, \citenamefont
  {Castin},\ and\ \citenamefont {Salomon}}]{khaykovich_formation_2002}%
  \BibitemOpen
  \bibfield  {author} {\bibinfo {author} {\bibfnamefont {L.}~\bibnamefont
  {Khaykovich}}, \bibinfo {author} {\bibfnamefont {F.}~\bibnamefont {Schreck}},
  \bibinfo {author} {\bibfnamefont {G.}~\bibnamefont {Ferrari}}, \bibinfo
  {author} {\bibfnamefont {T.}~\bibnamefont {Bourdel}}, \bibinfo {author}
  {\bibfnamefont {J.}~\bibnamefont {Cubizolles}}, \bibinfo {author}
  {\bibfnamefont {L.~D.}\ \bibnamefont {Carr}}, \bibinfo {author}
  {\bibfnamefont {Y.}~\bibnamefont {Castin}}, \ and\ \bibinfo {author}
  {\bibfnamefont {C.}~\bibnamefont {Salomon}},\ }\bibfield  {title}
  {{\selectlanguage {english}\enquote {\bibinfo {title} {Formation of a
  {Matter}-{Wave} {Bright} {Soliton}},}\ }}\href {\doibase
  10.1126/science.1071021} {\bibfield  {journal} {\bibinfo  {journal}
  {Science}\ }\textbf {\bibinfo {volume} {296}},\ \bibinfo {pages} {1290--1293}
  (\bibinfo {year} {2002})}\BibitemShut {NoStop}%
\bibitem [{\citenamefont {Strecker}\ \emph {et~al.}(2002)\citenamefont
  {Strecker}, \citenamefont {Partridge}, \citenamefont {Truscott},\ and\
  \citenamefont {Hulet}}]{strecker_formation_2002}%
  \BibitemOpen
  \bibfield  {author} {\bibinfo {author} {\bibfnamefont {Kevin~E.}\
  \bibnamefont {Strecker}}, \bibinfo {author} {\bibfnamefont {Guthrie~B.}\
  \bibnamefont {Partridge}}, \bibinfo {author} {\bibfnamefont {Andrew~G.}\
  \bibnamefont {Truscott}}, \ and\ \bibinfo {author} {\bibfnamefont
  {Randall~G.}\ \bibnamefont {Hulet}},\ }\bibfield  {title} {{\selectlanguage
  {english}\enquote {\bibinfo {title} {Formation and propagation of matter-wave
  soliton trains},}\ }}\href {\doibase 10.1038/nature747} {\bibfield  {journal}
  {\bibinfo  {journal} {Nature}\ }\textbf {\bibinfo {volume} {417}},\ \bibinfo
  {pages} {150--153} (\bibinfo {year} {2002})}\BibitemShut {NoStop}%
\bibitem [{\citenamefont {Cornish}\ \emph {et~al.}(2006)\citenamefont
  {Cornish}, \citenamefont {Thompson},\ and\ \citenamefont
  {Wieman}}]{cornish_formation_2006}%
  \BibitemOpen
  \bibfield  {author} {\bibinfo {author} {\bibfnamefont {Simon~L.}\
  \bibnamefont {Cornish}}, \bibinfo {author} {\bibfnamefont {Sarah~T.}\
  \bibnamefont {Thompson}}, \ and\ \bibinfo {author} {\bibfnamefont {Carl~E.}\
  \bibnamefont {Wieman}},\ }\bibfield  {title} {\enquote {\bibinfo {title}
  {Formation of {Bright} {Matter}-{Wave} {Solitons} during the {Collapse} of
  {Attractive} {Bose}-{Einstein} {Condensates}},}\ }\href {\doibase
  10.1103/PhysRevLett.96.170401} {\bibfield  {journal} {\bibinfo  {journal}
  {Phys. Rev. Lett.}\ }\textbf {\bibinfo {volume} {96}},\ \bibinfo {pages}
  {170401} (\bibinfo {year} {2006})}\BibitemShut {NoStop}%
\bibitem [{Note2()}]{Note2}%
  \BibitemOpen
  \bibinfo {note} {R.G.~Hulet, private communication.}\BibitemShut {Stop}%
\bibitem [{Note3()}]{Note3}%
  \BibitemOpen
  \bibinfo {note} {We note that in principle the relative Hamiltonian for any
  of the $N\protect \tmspace -\thinmuskip {.1667em}=\protect \tmspace
  -\thinmuskip {.1667em}4$ families also could be realized with a
  non-interacting Bose-Einstein condensate in a spherically-symmetric harmonic
  trap sliced by six sheets of intense laser light, or even just three sheets
  to create a sector.}\BibitemShut {Stop}%
\bibitem [{\citenamefont {Deutsch}(1991)}]{deutsch_quantum_1991}%
  \BibitemOpen
  \bibfield  {author} {\bibinfo {author} {\bibfnamefont {J.~M.}\ \bibnamefont
  {Deutsch}},\ }\bibfield  {title} {\enquote {\bibinfo {title} {Quantum
  statistical mechanics in a closed system},}\ }\href {\doibase
  10.1103/PhysRevA.43.2046} {\bibfield  {journal} {\bibinfo  {journal} {Phys.
  Rev. A}\ }\textbf {\bibinfo {volume} {43}},\ \bibinfo {pages} {2046--2049}
  (\bibinfo {year} {1991})}\BibitemShut {NoStop}%
\bibitem [{\citenamefont {Srednicki}(1994)}]{srednicki_chaos_1994}%
  \BibitemOpen
  \bibfield  {author} {\bibinfo {author} {\bibfnamefont {Mark}\ \bibnamefont
  {Srednicki}},\ }\bibfield  {title} {\enquote {\bibinfo {title} {Chaos and
  quantum thermalization},}\ }\href {\doibase 10.1103/PhysRevE.50.888}
  {\bibfield  {journal} {\bibinfo  {journal} {Phys. Rev. E}\ }\textbf {\bibinfo
  {volume} {50}},\ \bibinfo {pages} {888--901} (\bibinfo {year}
  {1994})}\BibitemShut {NoStop}%
\bibitem [{\citenamefont {Zhang}\ \emph {et~al.}(2015)\citenamefont {Zhang},
  \citenamefont {Dunjko},\ and\ \citenamefont {Olshanii}}]{zhang_atom_2015}%
  \BibitemOpen
  \bibfield  {author} {\bibinfo {author} {\bibfnamefont {Z.}~\bibnamefont
  {Zhang}}, \bibinfo {author} {\bibfnamefont {V.}~\bibnamefont {Dunjko}}, \
  and\ \bibinfo {author} {\bibfnamefont {M.}~\bibnamefont {Olshanii}},\
  }\bibfield  {title} {{\selectlanguage {english}\enquote {\bibinfo {title}
  {Atom transistor from the point of view of nonequilibrium dynamics},}\
  }}\href {\doibase 10.1088/1367-2630/17/12/125008} {\bibfield  {journal}
  {\bibinfo  {journal} {New J. Phys.}\ }\textbf {\bibinfo {volume} {17}},\
  \bibinfo {pages} {125008} (\bibinfo {year} {2015})}\BibitemShut {NoStop}%
\bibitem [{\citenamefont {Yurovsky}\ and\ \citenamefont
  {Olshanii}(2011)}]{yurovsky_memory_2011}%
  \BibitemOpen
  \bibfield  {author} {\bibinfo {author} {\bibfnamefont {V.~A.}\ \bibnamefont
  {Yurovsky}}\ and\ \bibinfo {author} {\bibfnamefont {M.}~\bibnamefont
  {Olshanii}},\ }\bibfield  {title} {\enquote {\bibinfo {title} {Memory of the
  {Initial} {Conditions} in an {Incompletely} {Chaotic} {Quantum} {System}:
  {Universal} {Predictions} with {Application} to {Cold} {Atoms}},}\ }\href
  {\doibase 10.1103/PhysRevLett.106.025303} {\bibfield  {journal} {\bibinfo
  {journal} {Phys. Rev. Lett.}\ }\textbf {\bibinfo {volume} {106}},\ \bibinfo
  {pages} {025303} (\bibinfo {year} {2011})}\BibitemShut {NoStop}%
\bibitem [{\citenamefont {Friedman}\ \emph {et~al.}(2001)\citenamefont
  {Friedman}, \citenamefont {Kaplan}, \citenamefont {Carasso},\ and\
  \citenamefont {Davidson}}]{friedman_observation_2001}%
  \BibitemOpen
  \bibfield  {author} {\bibinfo {author} {\bibfnamefont {Nir}\ \bibnamefont
  {Friedman}}, \bibinfo {author} {\bibfnamefont {Ariel}\ \bibnamefont
  {Kaplan}}, \bibinfo {author} {\bibfnamefont {Dina}\ \bibnamefont {Carasso}},
  \ and\ \bibinfo {author} {\bibfnamefont {Nir}\ \bibnamefont {Davidson}},\
  }\bibfield  {title} {\enquote {\bibinfo {title} {Observation of {Chaotic} and
  {Regular} {Dynamics} in {Atom}-{Optics} {Billiards}},}\ }\href {\doibase
  10.1103/PhysRevLett.86.1518} {\bibfield  {journal} {\bibinfo  {journal}
  {Phys. Rev. Lett.}\ }\textbf {\bibinfo {volume} {86}},\ \bibinfo {pages}
  {1518--1521} (\bibinfo {year} {2001})}\BibitemShut {NoStop}%
\bibitem [{\citenamefont {Sen}(1996)}]{sen_multispecies_1996}%
  \BibitemOpen
  \bibfield  {author} {\bibinfo {author} {\bibfnamefont {Diptiman}\
  \bibnamefont {Sen}},\ }\bibfield  {title} {\enquote {\bibinfo {title} {A
  multispecies {Calogero}-{Sutherland} model},}\ }\href {\doibase
  10.1016/0550-3213(96)00420-8} {\bibfield  {journal} {\bibinfo  {journal}
  {Nuc. Phys. B}\ }\textbf {\bibinfo {volume} {479}},\ \bibinfo {pages}
  {554--574} (\bibinfo {year} {1996})}\BibitemShut {NoStop}%
\bibitem [{\citenamefont {Werner}\ and\ \citenamefont
  {Castin}(2006)}]{werner_unitary_2006}%
  \BibitemOpen
  \bibfield  {author} {\bibinfo {author} {\bibfnamefont {F\'{e}lix}\
  \bibnamefont {Werner}}\ and\ \bibinfo {author} {\bibfnamefont {Yvan}\
  \bibnamefont {Castin}},\ }\bibfield  {title} {\enquote {\bibinfo {title}
  {Unitary gas in an isotropic harmonic trap: {Symmetry} properties and
  applications},}\ }\href {\doibase 10.1103/PhysRevA.74.053604} {\bibfield
  {journal} {\bibinfo  {journal} {Phys. Rev. A}\ }\textbf {\bibinfo {volume}
  {74}},\ \bibinfo {pages} {053604} (\bibinfo {year} {2006})}\BibitemShut
  {NoStop}%
\bibitem [{\citenamefont {Hamermesh}\ and\ \citenamefont
  {Physics}(1989)}]{hamermesh_group_1989}%
  \BibitemOpen
  \bibfield  {author} {\bibinfo {author} {\bibfnamefont {Morton}\ \bibnamefont
  {Hamermesh}}\ and\ \bibinfo {author} {\bibnamefont {Physics}},\ }\href@noop
  {} {{\selectlanguage {English}\emph {\bibinfo {title} {Group {Theory} and
  {Its} {Application} to {Physical} {Problems}}}}},\ \bibinfo {edition}
  {reprint edition}\ ed.\ (\bibinfo  {publisher} {Dover Publications},\
  \bibinfo {address} {New York},\ \bibinfo {year} {1989})\BibitemShut {NoStop}%
\bibitem [{\citenamefont {Wybourne}(1974)}]{wybourne_classical_1974}%
  \BibitemOpen
  \bibfield  {author} {\bibinfo {author} {\bibfnamefont {Brian~G.}\
  \bibnamefont {Wybourne}},\ }\href@noop {} {{\selectlanguage {English}\emph
  {\bibinfo {title} {Classical {Groups} for {Physicists}}}}},\ \bibinfo
  {edition} {first printing edition}\ ed.\ (\bibinfo  {publisher} {John Wiley
  \& Sons Inc},\ \bibinfo {address} {New York},\ \bibinfo {year}
  {1974})\BibitemShut {NoStop}%
\bibitem [{\citenamefont {Harmer}(2007)}]{harmer_spectra_2007}%
  \BibitemOpen
  \bibfield  {author} {\bibinfo {author} {\bibfnamefont {Mark}\ \bibnamefont
  {Harmer}},\ }\bibfield  {title} {\enquote {\bibinfo {title} {The spectra of
  the spherical and euclidean triangle groups},}\ }\href {\doibase
  10.1017/S1446788708000281} {\bibfield  {journal} {\bibinfo  {journal}
  {Journal of the Australian Mathematical Society}\ }\textbf {\bibinfo {volume}
  {84}},\ \bibinfo {pages} {217--227} (\bibinfo {year} {2007})}\BibitemShut
  {NoStop}%
\bibitem [{\citenamefont {Rittenhouse}\ \emph {et~al.}(2011)\citenamefont
  {Rittenhouse}, \citenamefont {{von Stecher}}, \citenamefont {D'Incao},
  \citenamefont {Mehta},\ and\ \citenamefont
  {Greene}}]{rittenhouse_hyperspherical_2011}%
  \BibitemOpen
  \bibfield  {author} {\bibinfo {author} {\bibfnamefont {S.~T.}\ \bibnamefont
  {Rittenhouse}}, \bibinfo {author} {\bibfnamefont {J.}~\bibnamefont {{von
  Stecher}}}, \bibinfo {author} {\bibfnamefont {J.~P.}\ \bibnamefont
  {D'Incao}}, \bibinfo {author} {\bibfnamefont {N.~P.}\ \bibnamefont {Mehta}},
  \ and\ \bibinfo {author} {\bibfnamefont {C.~H.}\ \bibnamefont {Greene}},\
  }\bibfield  {title} {{\selectlanguage {english}\enquote {\bibinfo {title}
  {The hyperspherical four-fermion problem},}\ }}\href {\doibase
  10.1088/0953-4075/44/17/172001} {\bibfield  {journal} {\bibinfo  {journal}
  {J. Phys. B: At. Mol. Opt. Phys.}\ }\textbf {\bibinfo {volume} {44}},\
  \bibinfo {pages} {172001} (\bibinfo {year} {2011})}\BibitemShut {NoStop}%
\end{thebibliography}
\end{document}